\documentclass[draftcls,onecolumn,12pt,journal,letter]{IEEEtran}  

\usepackage{epsfig,graphics}
\usepackage{subfigure}
\usepackage{amssymb,amsmath,amsfonts}
\usepackage[usenames]{color}
\usepackage{bbm}
\usepackage{psfrag}
\usepackage{cite}
\usepackage{latexsym}
\usepackage{amsmath}
\usepackage{multirow}
\usepackage{color}
\usepackage{booktabs}  
\usepackage{mathtools}

\IEEEoverridecommandlockouts

\newcommand{\tz}[1]{{\color{blue}#1}}

\title{Time- and Frequency-Varying $K$-Factor of Non-Stationary Vehicular Channels for Safety Relevant Scenarios}

\author{\IEEEauthorblockN{Laura Bernad\'o, {\it Member, IEEE,} Thomas Zemen${^*}$, {\it Senior Member, IEEE,} Fredrik Tufvesson, {\it Senior Member, IEEE,} Andreas F.{} Molisch, {\it Fellow, IEEE,} Christoph F.{} Mecklenbr\"auker, {\it Senior Member, IEEE}}
\thanks{
Manuscript received yyyyyyy.2014. L.~Bernad\'o and T.~Zemen ({\bf corresponding author}) are with Forschungszentrum Telekommunikation Wien (FTW), Vienna, Austria (e-mail: bernado@ftw.at, {\bf thomas.zemen@ftw.at}). F.~Tufvesson is with the Department of Electrical and Information Technology, Lund University, Lund, Sweden (e-mail: Fredrik.Tufvesson@eit.lth.se). C.~F.~Mecklenbr\"auker is with the Institut of Telecommunications, Technische Universit\"at Wien, Vienna, Austria (e-mail: cfm@nt.tuwien.ac.at). A.~F.~Molisch is with the Department of Electrical Engineering, University of Southern California, Los Angeles, CA, USA (e-mail: molisch@usc.edu).

This research was supported by the project NOWIRE funded by the Vienna Science and Technology Fund (WWTF), the project NFN SISE (S10607) funded by the Austrian Science Fund (FWF) as well as the strategic FTW project I-0. FTW is supported by the Austrian Government and the City of Vienna within the competence center program COMET. The research work described in this paper was carried out in cooperation with the COST Action IC1004 on Cooperative Radio Communications for Green Smart Environments. Part of this research has been presented in the \emph{IEEE 21st International Symposium on Personal Indoor and Mobile Radio Communications (PIMRC), 2010}.
}}

\begin{document}

\maketitle

\begin{abstract}
Vehicular communication channels are characterized by a non-stationary time- and frequency-selective fading process due to fast changes in the environment. We characterize the distribution of the envelope of the first delay bin in vehicle-to-vehicle channels by means of its Rician $K$-factor. We analyze the time-frequency variability of this channel parameter using vehicular channel measurements at 5.6 GHz with a bandwidth of 240 MHz for safety-relevant scenarios in intelligent transportation systems (ITS). This data enables a frequency-variability analysis from an IEEE 802.11p system point of view, which uses 10 MHz channels. We show that the small-scale fading of the envelope of the first delay bin is Rician distributed with a varying $K$-factor. The later delay bins are Rayleigh distributed. We demonstrate that the $K$-factor cannot be assumed to be constant in time and frequency. The causes of these variations are the frequency-varying antenna radiation patterns as well as the time-varying number of active scatterers, and the effects of vegetation. We also present a simple but accurate bi-modal Gaussian mixture model, that allows to capture the $K$-factor variability in time for safety-relevant ITS scenarios.
\end{abstract}

\section{Introduction}
Intelligent transportation systems (ITS) have gained much interest in the last years. They have the potential to strongly reduce the rate of accidents and environmental pollution with the help of reliable wireless vehicle-to-vehicle (V2V) and vehicle-to-infrastructure (V2I) communications. 

New receiver algorithms for vehicular communications are first evaluated using numeric link level simulations which rely on accurate channel models. Key factors influencing the wave propagation in vehicular channels are the low position of the antenna on the rooftop of the vehicle, the larger number of metallic objects close to the communication link, and the high mobility of the transmitter (TX) and the receiver (RX). The combination of all these aspects, as in vehicular communications, give rise to a non-stationary fading process. Therefore, it is necessary to study and characterize the small-scale fading statistics in vehicular channels, see e.g. \cite{Matz2005, Wilink2008, Bernado2008, Renaudin2010, Mecklenbraeuker2011, Molisch2009, Maurer2002, Acosta2004, Bernado2014}, since non-stationary vehicular channel properties have a large impact on the performance of a communication system \cite{Zemen12, Zemen12a}. 

In this paper we particularly focus on the characteristic of the Rician $K$-factor of non-stationary vehicular channels. The $K$-factor 
\begin{equation}
K =10 \text{log}_{10}(r^2/2\sigma^2) \,\text{[dB]} 
\label{eq:Kfactor}
\end{equation}
is defined as the ratio of the energy of the specular part $r^2$ and the diffuse part $2\sigma^2$ of the received signal \cite{Rice1948,Durgin2002}. Often, the specular part consists only of the line of sight (LOS) component, but specular components can also stem from flat good reflecting surfaces (e.g. traffic signs) where the impinging wave is reflected into a single direction.

The use of the Rician $K$-factor is widespread for describing the small-scale fading characteristics. Therefore the $K$-factor is frequently used as input parameter in numerical simulations to test receiver algorithms, coding schemes, or cooperative strategies \cite{Castiglione2010, Lin2009}. Often the $K$-factor is assumed to be deterministic and is used as constant value for these simulations. However, due to the non-stationary nature of vehicular communication channels this assumption does not hold true, as we will demonstrate in this paper.

In the literature there are only few contributions that characterize the $K$-factor in vehicular environments at $5\,$GHz. In \cite{Maurer2002} and \cite{Karedal2009b} the authors show by means of channel measurements that the envelope of the first delay bin is Rician distributed, but they do not give concrete values. $K$-factors between $-15\,$dB and $15\,$dB are reported in \cite{Thiele2006} at $5.2\,$GHz for a measurement bandwidth of $120\,$MHz, with an average value of $2.75\,$dB. Results in the $5\,$GHz band are also presented in \cite{Renaudin2008}, where the $K$-factor is obtained for a $60\,$MHz bandwidth system and a value of about $3.5\,$dB is reported for different measurement conditions.  

All these contributions do not analyze the variability of the small scale fading statistics and only few investigations are available that consider the actual randomness of the $K$-factor. In \cite{Delibasic2012} an analytic model is discussed that takes the randomness of the $K$-factor into account, but no validation with channel measurements is presented. In \cite{Messier2009, Greenstein2009} the non-constant $K$-factor is analyzed based on measurement data centered at $1.9\,$GHz considering mobility in a cellular setting. In \cite{Messier2009} an analytical model is presented for a dense urban environment using a bandwidth of $1.23\,$MHz; and in \cite{Greenstein2009} the authors examine the time and frequency variability of the $K$-factor, and they show a fairly constant $K$-factor in the frequency domain within the measurement bandwidth of $9\,$MHz. 

A first preliminary analysis of the time- and frequency-varying $K$-factor for vehicular communications only for a single rural scenario is presented in \cite{Bernado2010b}. 

\subsection*{Contributions of the Paper:}
\begin{itemize}
\item We focus on the characterization of the small-scale fading statistics for non-stationary V2V channels in a large set of different safety-relevant scenarios for ITS using channel measurement data obtained in the DRIVEWAY'09 measurement campaign \cite{Paier2010}. \tz{To the best of our knowledge no other publication is available dealing with these important scenarios for ITS.}
\item We investigate the variation of the $K$-factor of non-stationary V2V channels in time and frequency. The analysis is done from an IEEE 802.11p system point of view, i.e. the parameters for the analysis are chosen such that they are standard compliant and therefore reliable results for ITS systems can be derived.
\item A simple but accurate bi-modal Gaussian mixture distribution is proposed to model the empiric $K$-factor distribution for each safety-relevant scenario. This approach is related to the one used for the root mean square (RMS)-delay and RMS-Doppler spread modeling in the companion paper \cite{Bernado2014} for the same scenario set.
\end{itemize}

\subsection*{Organization of the Paper:}
The channel measurements used for carrying out the investigation are described in Sec. \ref{sec:measurements}. The methodology for estimating the $K$-factor is presented in Sec. \ref{sec:Kestim}. The results on the time-frequency variability of the $K$-factor are discussed in Sec. \ref{sec:empiricalresults} for a specifically chosen illustrative exemplary measurement. A statistical modeling of the $K$-factor is proposed in Sec. \ref{sec:modeling}, and the results are discussed in detail. We conclude the paper in Sec. \ref{sec:conclusions}.

\section{Measurement Data Description}
\label{sec:measurements}

For the analysis of the Rician $K$-factor variability of non-stationary vehicular channels in time and in frequency we use vehicular radio channel measurements collected in the DRIVEWAY'09 measurement campaign \cite{Paier2010}. The channel measurements were conducted at a carrier frequency of $f_\text{C}=5.6\,$GHz with a measurement bandwidth of $B=240\,$MHz in $N=769$ frequency bins, using the RUSK-Lund channel sounder, based on the switched array sounding principle \cite{Thomae2000}. The measurements were performed for the following six safety-relevant scenarios in ITS:
\begin{enumerate}
\item road-crossing, in urban and suburban environments; 
\item general LOS obstruction on the highway; 
\item merging lanes in a rural environment; 
\item traffic congestion in two different situations, slow traffic and approaching a traffic jam; 
\item in-tunnel; and 
\item {on-bridge}. 
\end{enumerate}
More details regarding the measurement scenarios can be found in \cite{Bernado2014, Bernado12b}. 

The vehicles containing the transmitter (TX) and receiver (RX) parts of the channel sounder were equipped with a 4 element linear antenna array. Each antenna element has a main lobe that covers one of the $4$ main propagation directions. The transmit and receive antenna indices $n_\text{RX},n_\text{TX}\in\{1,2,3,4\}$ correspond to the main radiation directions $\{\text{left},\text{back}, \text{front}, \text{right}\}$ \cite{Klemp2010}, see Fig. \ref{fig:scenario}. With this $4\times 4$ antenna configuration we obtain 16 spatial links between TX and RX. The mapping between the link index $\ell$ and the TX and RX antenna indices is given by 
\begin{equation}
\ell= 4 (n_\text{TX}-1) + (5-n_\text{RX})\,,
\end{equation}
see also \cite[Fig. 5.5 and Tab. 5.2]{Bernado2012thesis}.

For the $K$-factor analysis we use the sampled time-varying frequency response, defined as
\begin{equation}
H[m,b,\ell] \coloneqq H(m t_\text{s}, b f_\text{s},\ell),
\label{eq:CFR}
\end{equation}
where $m$, $b$, and $\ell$ represent discrete time, discrete frequency and the link number while $t_\text{s}$ and $f_\text{s}$ denote the time and frequency resolution, respectively. 

The space variability of the $K$-factor was already demonstrated in \cite{Bernado2010b}, therefore we center the analysis performed in the current work on the time-frequency-varying $K$-factor, thus concentrating on a single link $\ell=10$, corresponding to the \emph{front-front} antenna configuration, between the antenna pair $n_\text{TX}=3$ and $n_\text{RX}=3$. We therefore drop the link index $\ell$ in all equations for the rest of the manuscript.

The measurement parameters used for the $K$-factor investigation are summarized in Tab. \ref{tab:params}.
\begin{table}
\begin{center}
\caption{Channel measurement parameters}
\label{tab:params}
\begin{tabular}{@{}lr@{}} \toprule
parameter             &   value  \\
\midrule
channel            & $4\times4$ MIMO \\

carrier frequency         & $f_\text{C}=5.6$\,GHz    \\

measurement bandwidth       & $B=240$\,MHz\\

transmit power          & $27$\,dBm     \\

sampling sequence length & $3.2\,\mu$s\\

frequency bins          & $N=769$\\

time resolution    & $t_\text{s}=307.2$\,$\mu$s \\
frequency resolution & $f_\text{s}=B/N=312.1\,$kHz\\

recording time          & $10 \ldots 20$\,s \\
\bottomrule
\end{tabular}
\end{center}
\end{table}

\section{$K$-factor Estimation}
\label{sec:Kestim}

In order to conduct a time-frequency-varying $K$-factor analysis from an IEEE 802.11p system point of view with the available measurement data, we will proceed as follows:
\begin{itemize}
\item \emph{Time-variability:} For each scenario 3 to 15 measurement runs were recorded, each with a duration of $10$ or $20$ s. Hence, this allows to characterize the time variation of the $K$-factor.
\item \emph{Frequency-variability:} The IEEE 802.11p standard for vehicular communications utilizes a communication bandwidth of $B_c=10\,$MHz. Having $B=240\,$MHz measured bandwidth available, we can conduct a frequency-dependent investigation of the small scale fading over $Q=B/B_c=24$ sub-bands of $10\,$MHz each. The sampling rate in the delay domain is given by the $10$\,MHz bandwidth of each sub-band.
\end{itemize}

The investigation of the $K$-factor is conducted in the delay domain. For that purpose, we first apply an inverse discrete Fourier transform to the recorded channel transfer function in order to obtain the time-frequency dependent channel impulse response (CIR) for all Q frequency sub-bands
\begin{equation}
h[m,n;q]=\sum_{c=0}^{N_c-1}H[m,q(N_c-1)+c]W[m,c]\text{e}^{\text{j}2\pi c n / N_c}
\end{equation}
where $m\in\{0,\ldots,S-1\}$ denotes the time index, $n\in\{0,\ldots,N_c-1\}$ the delay index, $q\in\{0,\ldots,Q-1\}$ the frequency sub-band index, and $W$ is a Hanning window of length $N_c$. The values used in the measurements in this publication are $S=32500$ or $S=65000$ for a $10\,$s or $20\,$s measurement run respectively, $N_c=33$, and $Q=24$.

\subsection{Data Preprocessing}
\begin{figure}
\begin{center}
\includegraphics[width=0.90\columnwidth]{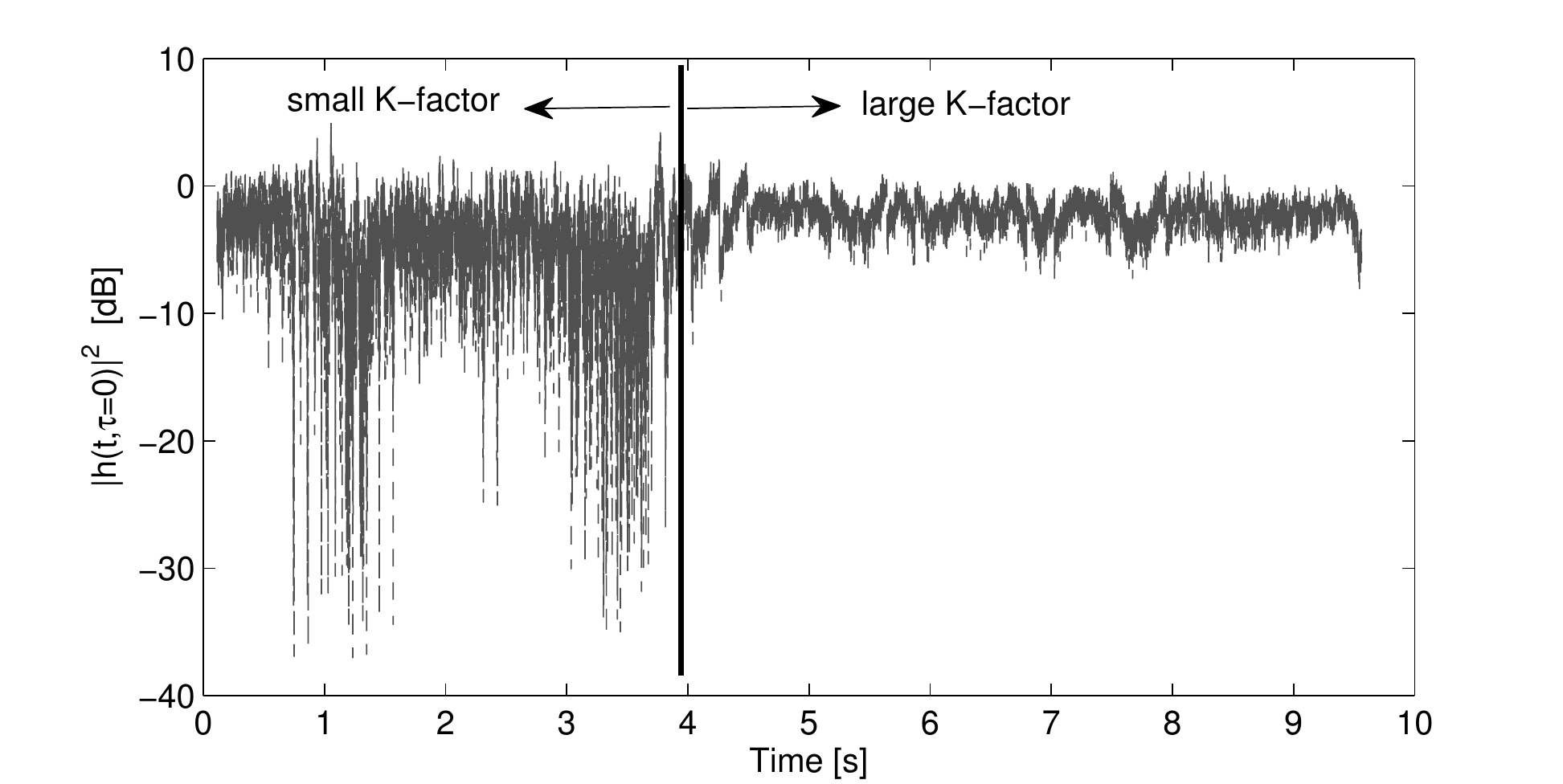}
\caption{Magnitude square of the envelope of the {CIR} of the first delay bin $|h'[m,n=0]|^2$ without large-scale fading.}
\label{fig:04:09}
\end{center}
\end{figure}

We characterize the distribution of the envelope of the first strong delay bin. For that, we need to first conduct some data preprocessing, which consists of two steps: 
\begin{enumerate}
\item Search for the delay bins corresponding to the first strong multipath component (MPC) in the CIR, and shift them to the origin ($\tau=0\,$ns). This is done on a per link, per frequency sub-band, and per time instance basis. We first perform a maximum search in data chunks of $S_\text{LS}$ time samples length, where $S_\text{LS}$ is the number of samples that will be used for removing the large scale fading effects. We then shift the CIR towards the origin by a number of delay bins corresponding to the mode where the maximum is placed within the chunk. This approach does not eliminate the small-scale fading effects.  
\item Remove the large-scale fading from the measured data by applying a moving average filter of length $S_\text{LS}$ samples in the time domain. We assume the large-scale fading to be stationary during $S_\text{LS}$ samples \cite{Clarke1968,Messier2009}. 
The large-scale fading changes slower than the small-scale fading effects, as a result, the averaging window $S_\text{LS}$ has to be considered at least as large as the number of samples used later for the $K$-factor estimation. For this reason, we set $S_\text{LS}>S_\text{K}$, where $S_\text{K}$ denotes the length of the observation window used for $K$-factor estimation.

The {CIR} without large-scale fading is calculated as
\begin{equation}
h'[m,n;q]=\frac{h[m,n;q]}{\sqrt{\varepsilon_h[m,n;q]}},
\label{eq:filterLS}
\end{equation}
where 
\begin{equation}
\varepsilon_h[m,n;q]=\frac{1}{S_\text{LS}}\sum_{m'=m-S_\text{LS}/2}^{m+S_\text{LS}/2-1}\sum_{n'=0}^{N-1}\left|h[m',n';q]\right|^2,
\label{eq:power}
\end{equation}

is the average power, $m\in\{S_\text{LS}/2, \ldots, S-1-S_\text{LS}/2\}$ represents \emph{absolute} time, and $m'\in\{m-S_\text{LS}/2, \ldots, m+S_\text{LS}/2-1\}$ denotes the \emph{local} time interval used to calculate the average power.

After large scale fading removal, the variance of a delay component is an indicator of its $K$-factor. When the received signal is composed of one specular paths its power does not vary. Whereas when the received signal is composed of additional diffuse paths, the superposition of them results in constructive or destructive interference, thus, the envelope varies significantly from one snapshot to the other. The more diffuse components the signal contains, the more variability, and also the smaller the $K$-factor.

As an example, we show in Fig. \ref{fig:04:09} the envelope of $h'[m,n;q]$ versus time $m$, for delay $n=0$, and sub-band $q=0$. In this figure we depict a setting where the $K$-factor is expected to be low up  to $4\,\text{s}$ and afterwards the $K$-factor is expected to be high, corresponding to periods where the variance of the envelope is large or small, respectively. This will be corroborated in the following sections.

\end{enumerate}

\begin{figure}[h]
\begin{center}
\includegraphics[width=0.90\columnwidth]{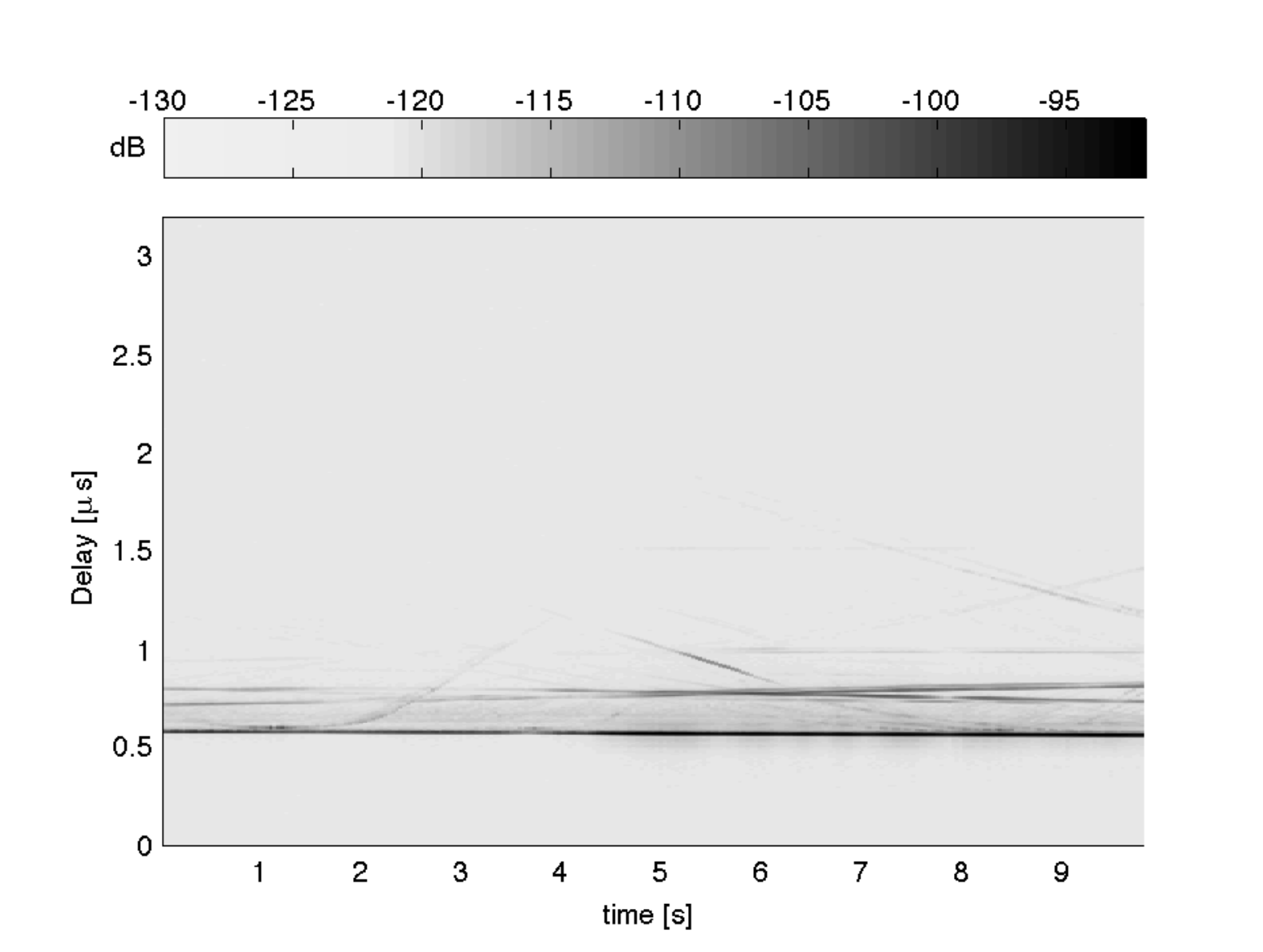}
\caption{Time-varying PDP for a \emph{{general LOS obstruction on highway}} scenario: convoy measurement with temporarily obstructed LOS (oLOS), constant velocities of $100$\,km/h ($27.8\,$m/s).}
\label{fig:GLO}%
\end{center}
\end{figure}

For conducting a first analysis of the $K$-factor, we use the \emph{general LOS obstruction} scenario on the \emph{highway} as an illustrative measurement example. Figure \ref{fig:GLO} shows the time-varying power delay profile (PDP) for this measurement. It will be useful for the reader to refer to this picture as a support for the $K$-factor results discussed in the following sections.
The {TX} and {RX} are driving in the same direction on the highway at around $100$\,km/h ($27.8\,$m/s) each, with big trucks driving in both directions beside them, and sometimes obstructing the {LOS} between the {TX} and the {RX} cars.

In this section we focus on the time-frequency variability of the $K$-factor using a single exemplary measurement from link $\ell=10$ between the antenna pair $n_\text{TX}=3$ and $n_\text{RX}=3$, as indicated previously. Later in Section \ref{sec:modeling} we perform a statistical analysis of all performed measurement runs in all scenarios.

\begin{figure}
\begin{center}
\includegraphics[width=0.90\columnwidth]{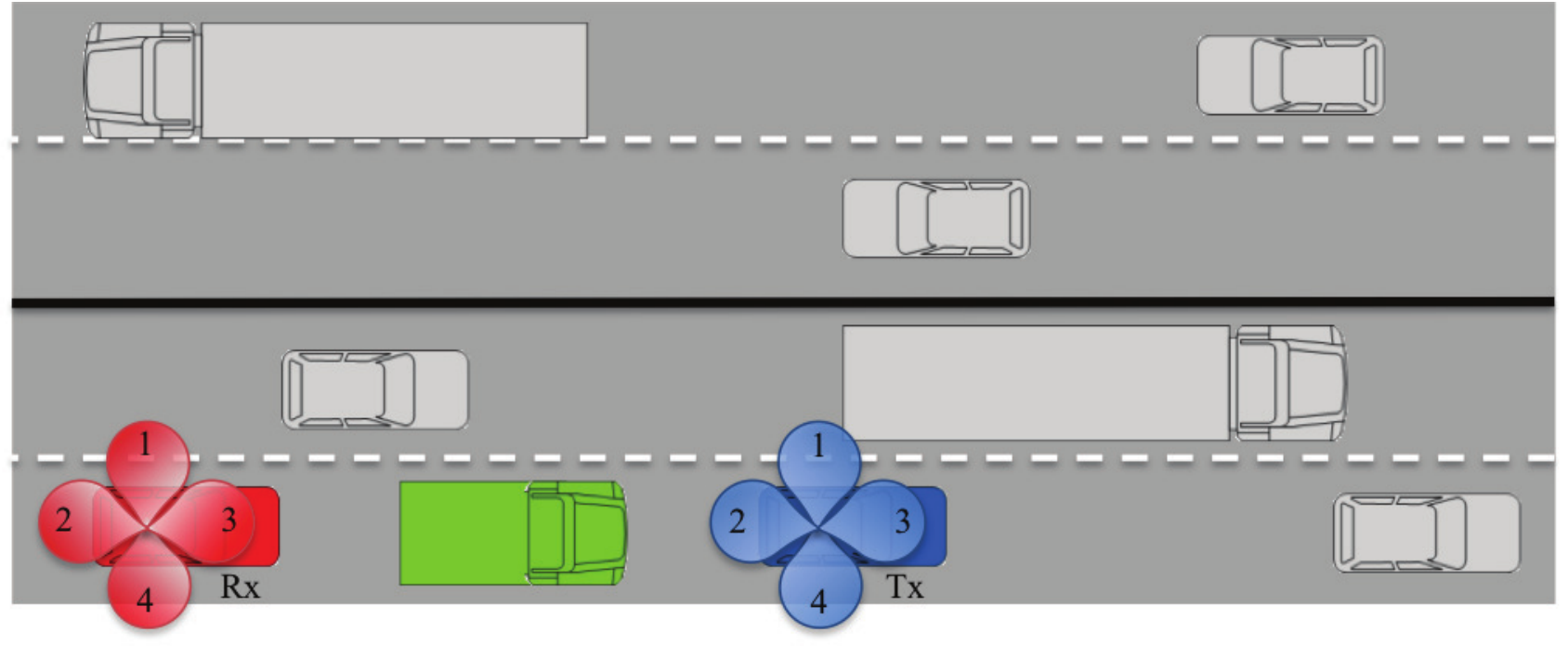}
\caption{Scenario layout - \emph{{general LOS obstruction on highway}}. In blue the {TX} car, in red the {RX} car. The green truck is the element intermittently obstructing the {LOS} during the measurement.}
\label{fig:scenario}
\end{center}
\end{figure}

\begin{figure}
\begin{center}
\subfigure[{TX} antenna $3$.]{\includegraphics[width=0.49\columnwidth]{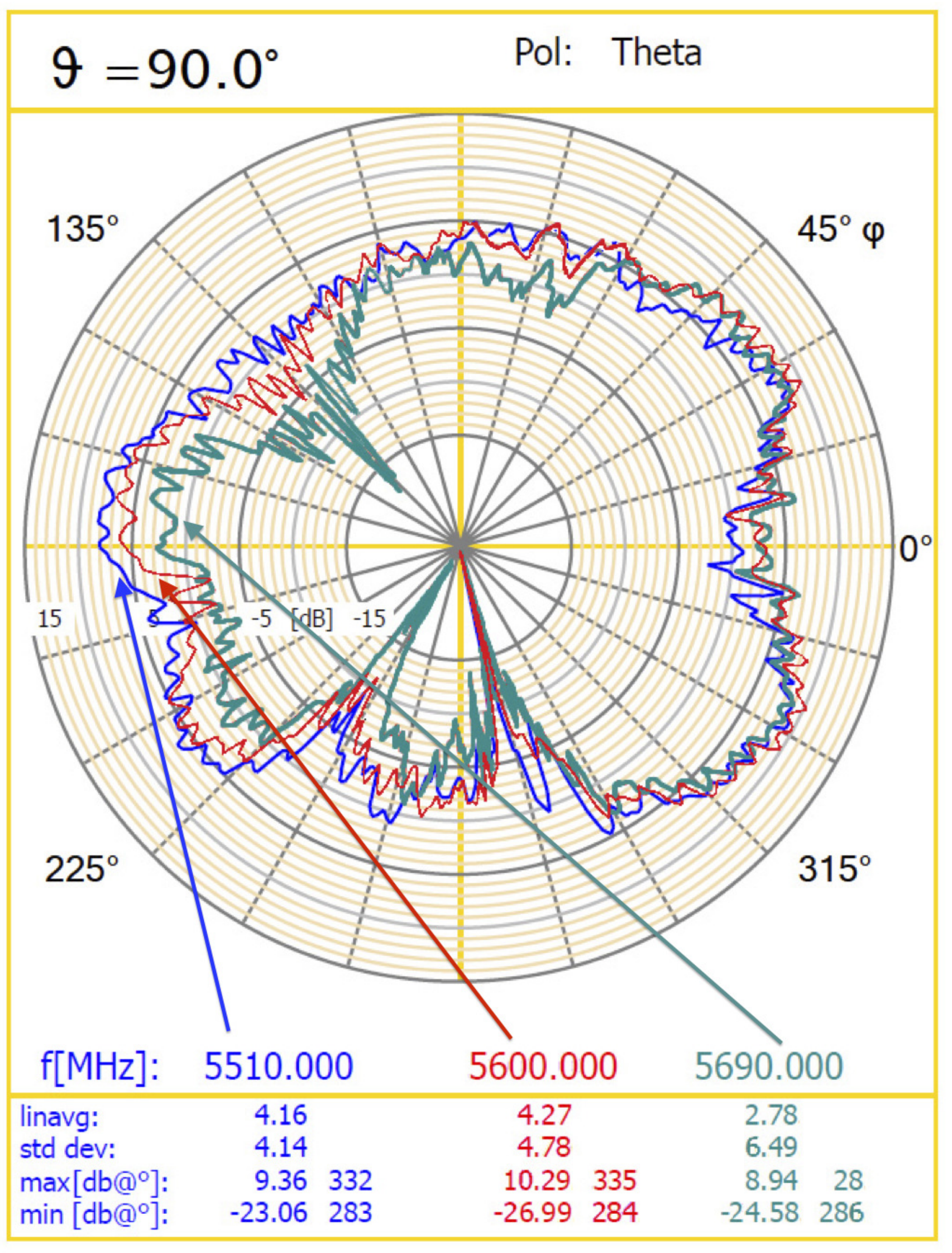}}
\subfigure[{RX} antenna $3$.]{\includegraphics[width=0.49\columnwidth]{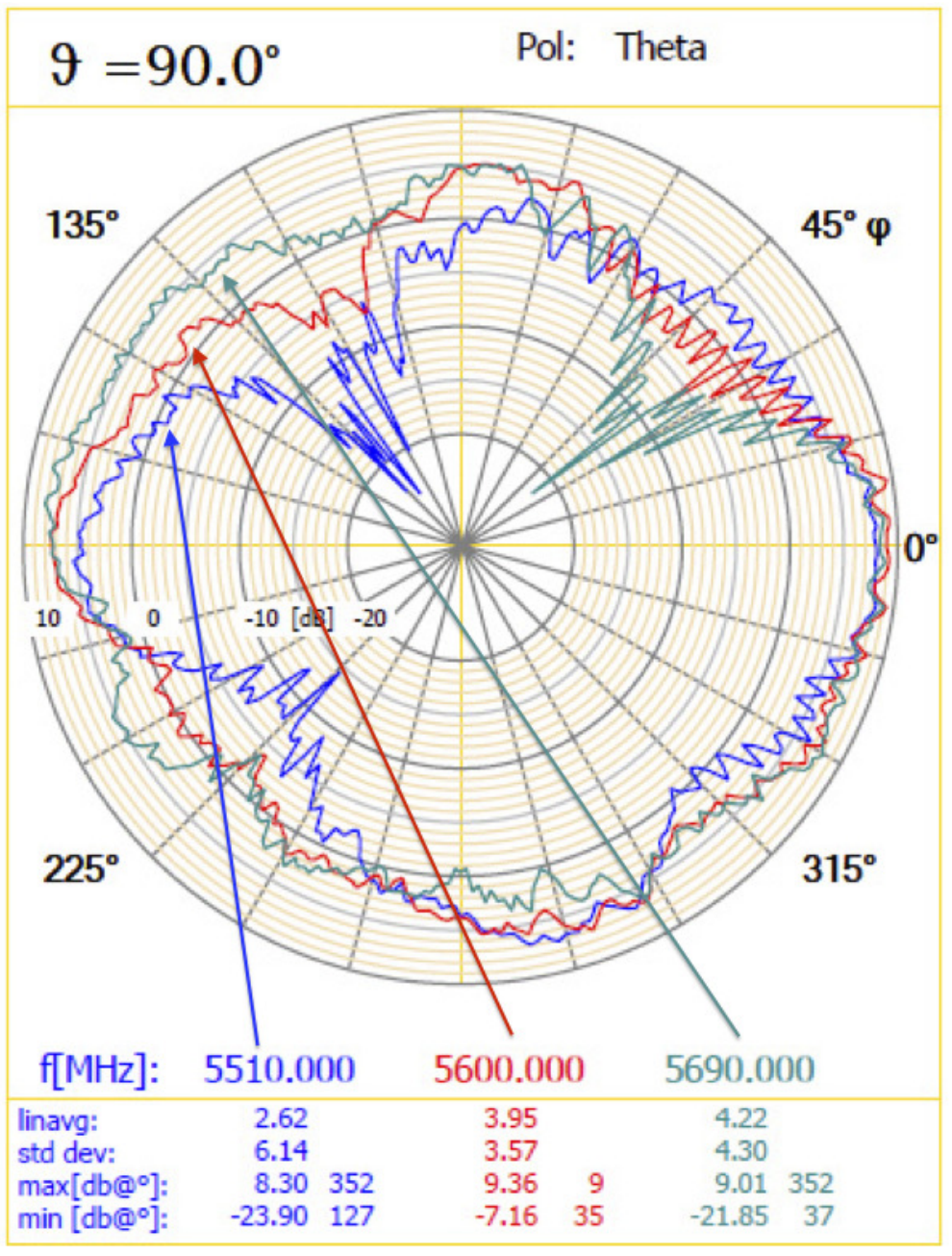}}
\caption{Radiation pattern of antenna element $3$, $\varphi = 0^{\circ}$ points into driving direction \cite{Bernado2011b}.}
\label{fig:04:15}
\end{center}
\end{figure}

Details on the radiation pattern of antenna element $n_\text{TX}=n_\text{RX}=3$ can be found in Fig. \ref{fig:04:15}, see also \cite{Bernado2010b}. Figure \ref{fig:scenario} shows an schematic view of the orientation of the radiation patterns for the TX and the RX.

\subsection{Sample Size Selection}
It is important to properly select the number of samples used for estimating the $K$-factor. The sample size $S_\text{K}$ has to be chosen such that the stationarity length of the process is not exceeded. On the other hand, a large number of samples is needed in order to obtain meaningful statistical results. Furthermore, the condition $S_\text{K} < S_\text{LS}$ must be fulfilled.

We choose the length $S_\text{K}$ such that it corresponds to $100\,\lambda$ travelled distance. Since the snapshot repetition time is fixed to $t_s=307.2\mu s$, the number of samples varies depending on the measured scenario, i.e. depending on the speed of the TX and RX. Table \ref{tab:SK} lists the average speed and the selected $S_K$ for the different scenarios \footnote{Due to unreliable GPS data, the window length $S_\text{K}$ is not adapted to the instantaneous velocity. Instead, we use average speed values to compute $S_{K}$.}. The number of samples used for removing the large-scale fading effects is then selected as $S_\text{LS}=S_\text{K}+2\Delta S$, being added $\Delta S=50$ samples at the beginning and end of the window $S_\text{K}$.

\begin{table}
\begin{center}
\caption{$S_\text{K}$: Number of samples used for estimating the $K$-factor}
\label{tab:SK}
\begin{tabular}{llclc} 
\toprule
Scenario	      						& Average speed       			& Number of samples $S_\text{K}$ \\
\midrule
\emph{road crossing}            			& $8.3$\,m/s ($30$\,km/h)		& $2100$\\
\emph{general LOS obstruction}            	& $27.8$\,m/s ($100$\,km/h) 		& $630$\\
\emph{merging lanes}            			& $22.2$\,m/s ($80$\,km/h)		& $790$\\
\emph{slow traffic}            			& $5.5$\,m/s ($20$\,km/h)		& $3100$\\
\emph{approaching traffic jam}   		& $16.7$\,m/s ($60$\,km/h)		& $1050$\\
\emph{in-tunnel}					& $25$\,m/s ($90$\,km/h)			& $700$\\
{\emph{on-bridge}}					& {$27.8$\,m/s ($100$\,km/h)}		& {$630$}\\
\bottomrule
\end{tabular}
\end{center}
\end{table}

\subsection{Envelope Distribution per Tap}
\label{se:EvelopeDistributionTap}
\begin{figure}
\begin{center}
{\includegraphics[width=1.1\columnwidth]{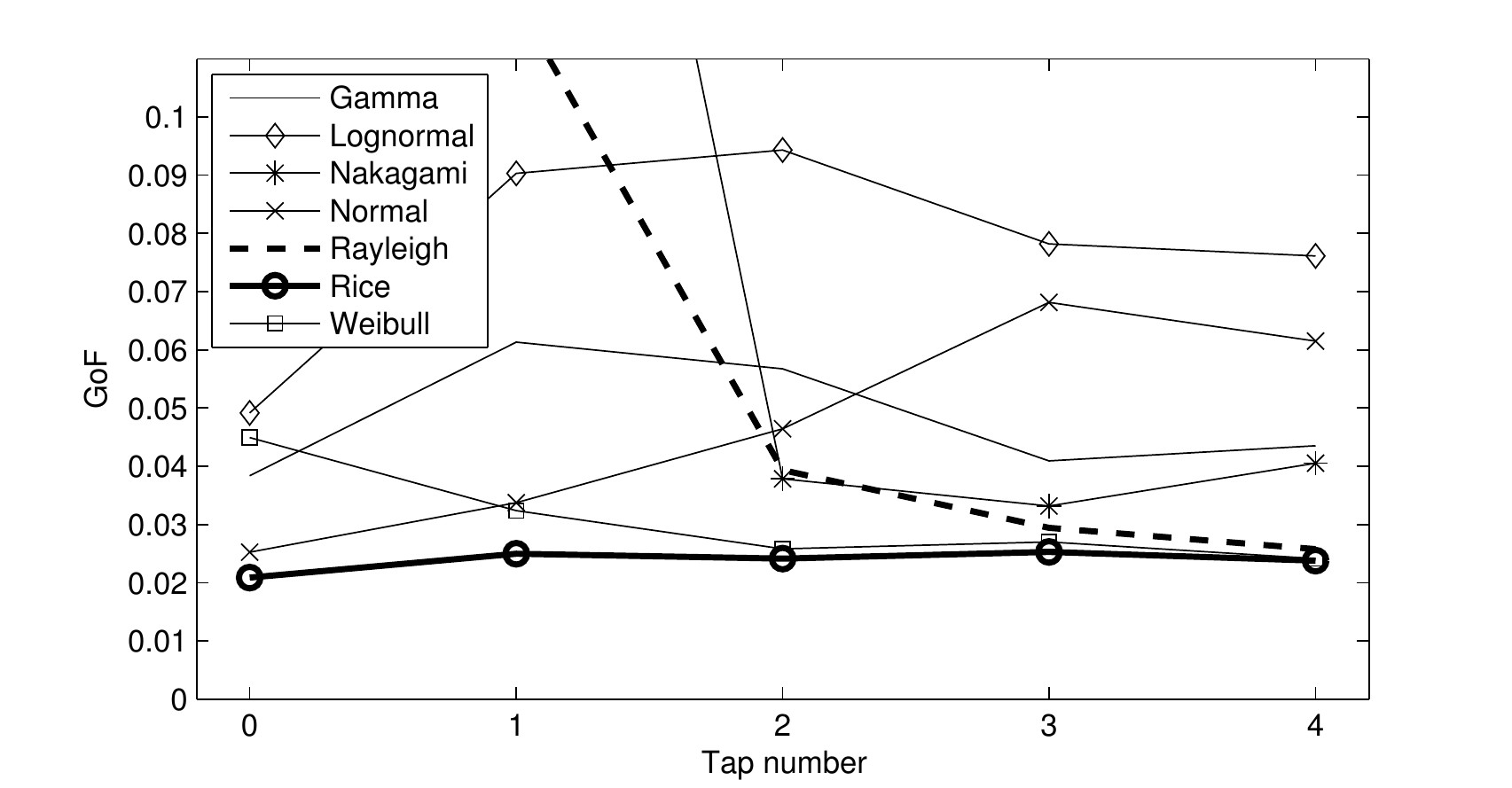}}
\caption{Goodness of fit for the first $5$ taps using different distributions to fit their envelope.}
\label{fig:GoF_envelope}
\end{center}
\end{figure}

Even though it is well accepted that the Rician distribution is the best for describing the small-scale fading effects, before further proceeding with the $K$-factor analysis, we evaluate the fitting of the envelope of the first $5$ delay bins to different distributions. In order to decide which distribution fits best, we apply the Kolmogorov-Smirnov (KS) test as a goodness-of-fit (GoF) indicator \cite{Massey1951}, calculated as
\begin{equation}
\text{GoF} =\sup_z|F_Z(z)-F_0(z)|\,,
\label{eq:GoF}
\end{equation}
where $\sup$ denotes the supremum, and $F_Z(z)$ is the empirical cumulative distribution function (CDF), and $F_0(z)$ is the analytical CDF, of the random variable $z$. We consider to achieve a good fit if the outcome of a {KS}-test is below $\epsilon$, which indicates that the maximum distance between empirical and fitted {CDF} is lower than $\epsilon$. Figure \ref{fig:GoF_envelope} shows the GoF result for the first $5$ taps for different distributions for the exemplary measurement run for link $\ell=10$, and $q=11$, corresponding to $f=5600\,$MHz. 
Different markers correspond to different distributions, the GoF plotted in the figure is the result of averaging the GoF values for $3$ time instances ($t=0.1\,\text{s, }t=4.8\,\text{s, }t=9.8\,\text{s}$). We corroborate from this figure, that the Rician distribution fits best the data, thus we carry on with the $K$-factor analysis.

\begin{figure}
\begin{center}
\subfigure[$t=0.1\,$s]{\includegraphics[width=0.7\columnwidth]{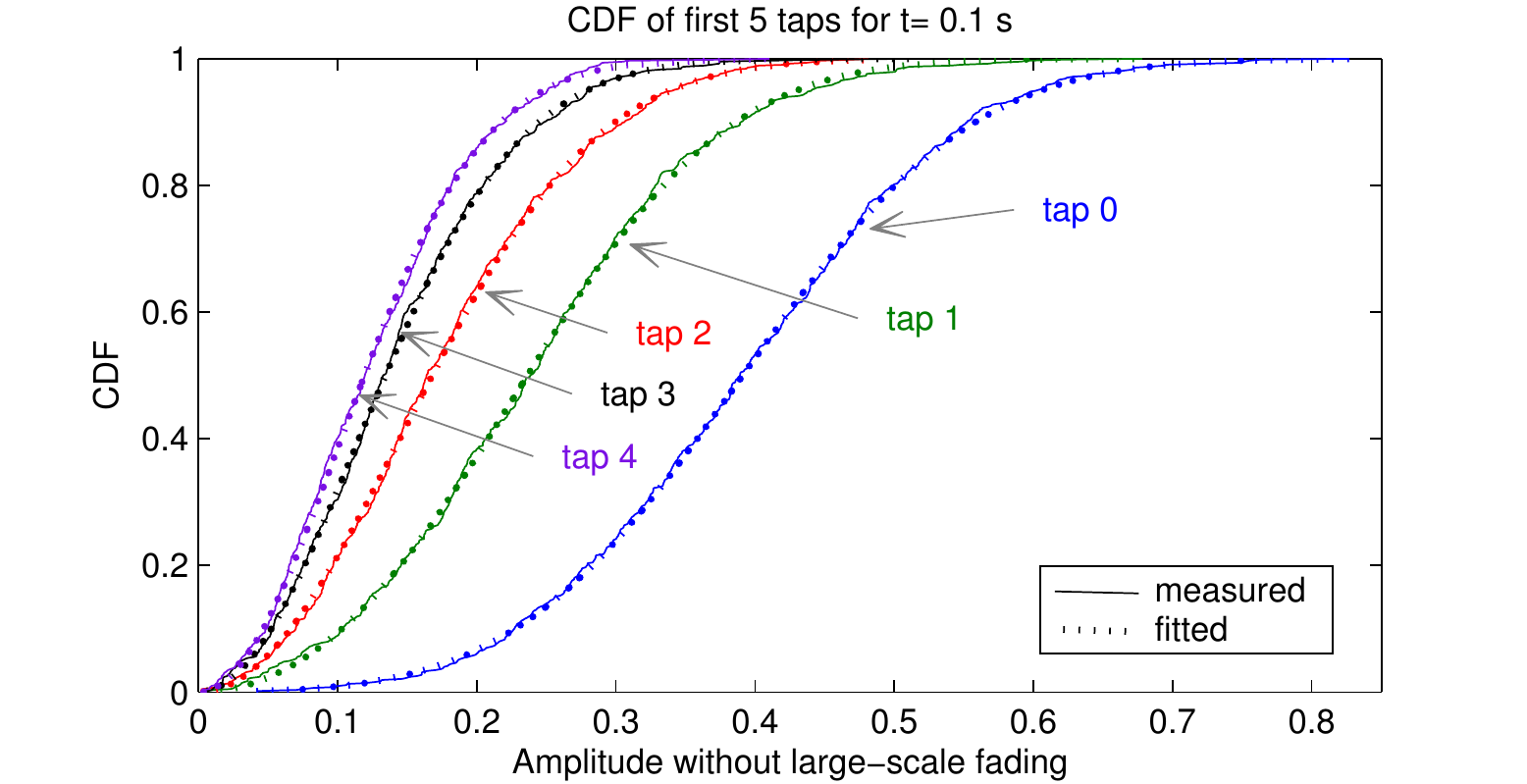}} 
\subfigure[$t=4.8\,$s]{\includegraphics[width=0.7\columnwidth]{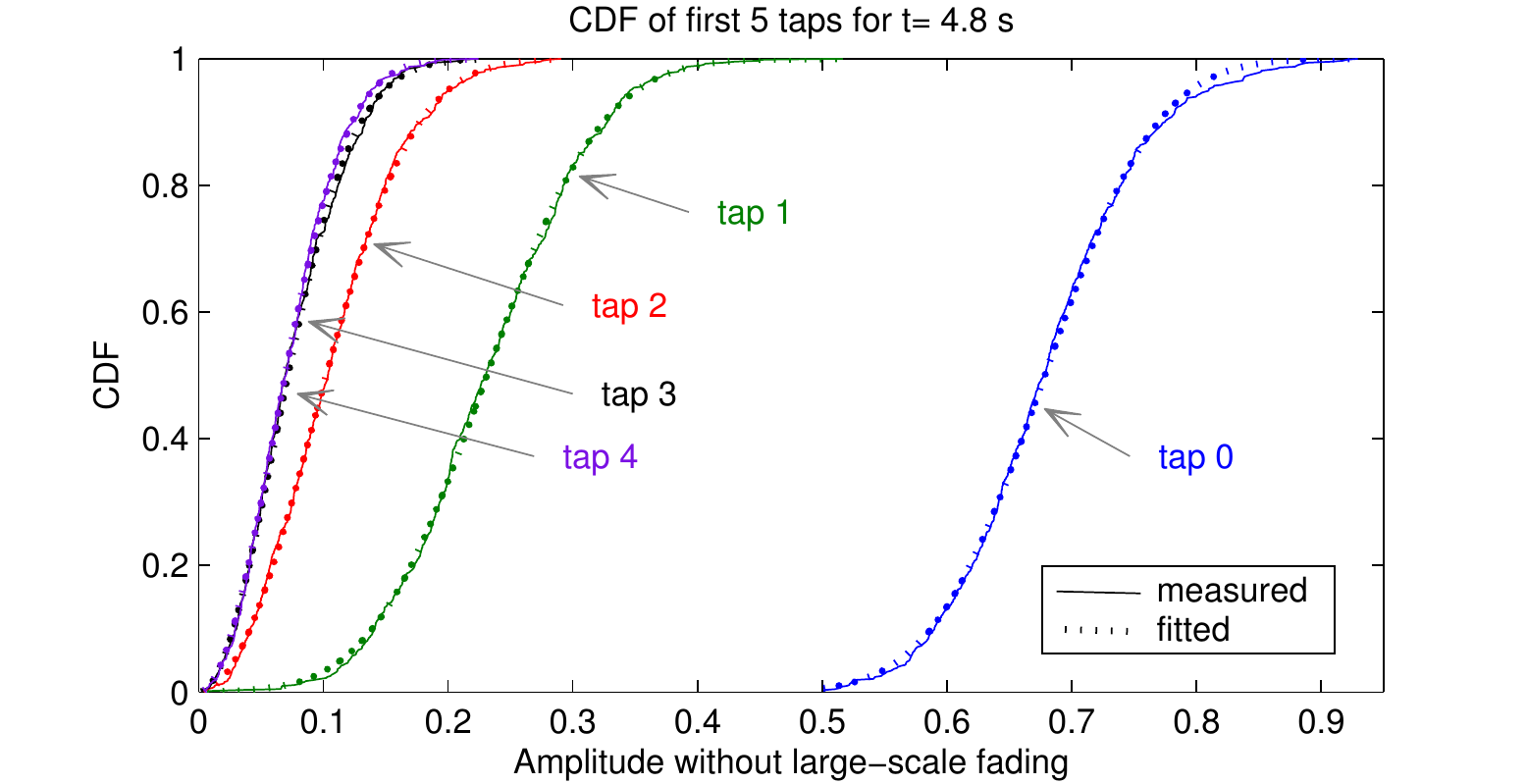}}
\subfigure[$t=9.8\,$s]{\includegraphics[width=0.7\columnwidth]{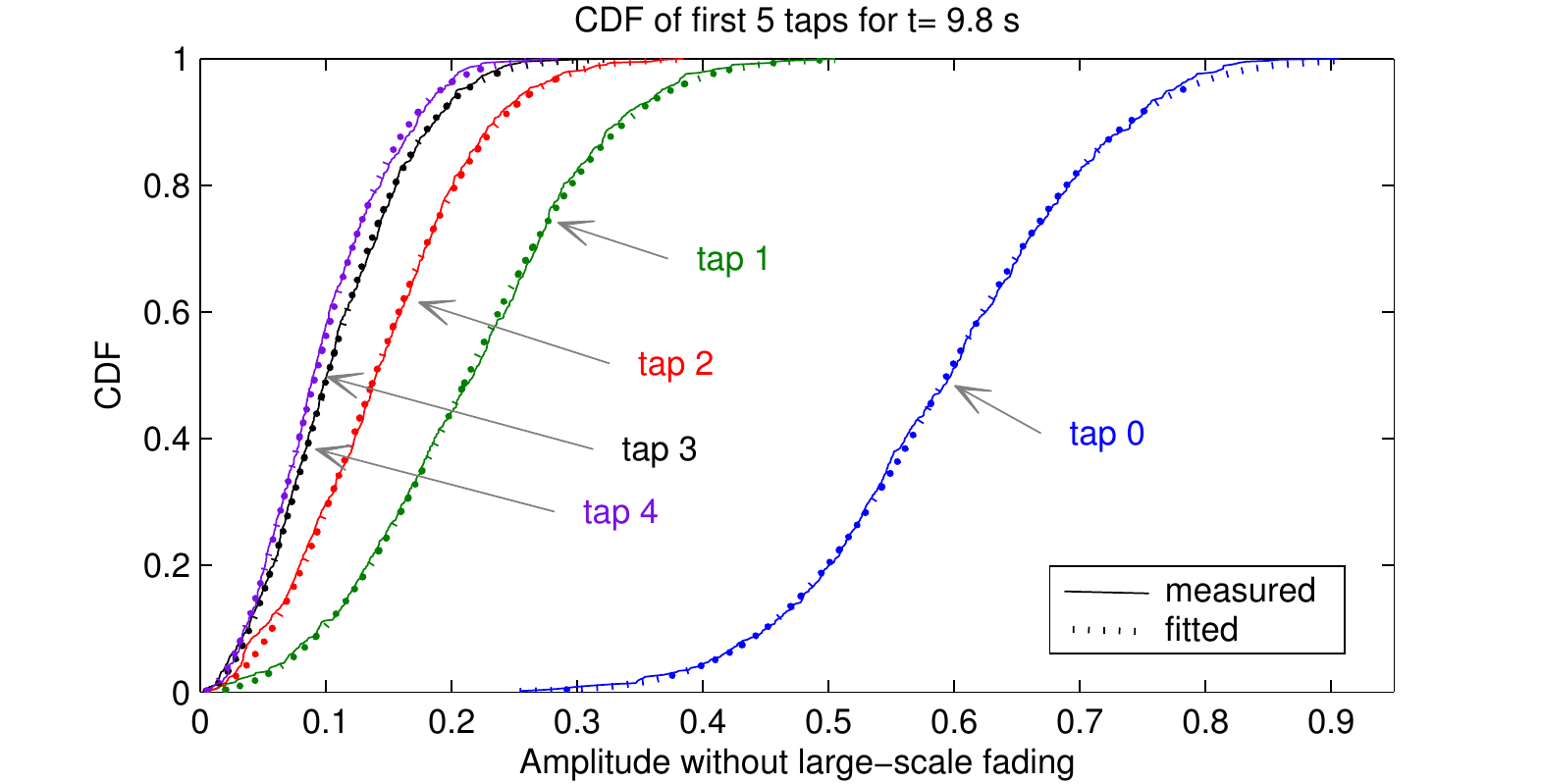}}
\caption{Empirical and fitted {CDF} for delay bins $n\in\{0,\ldots,4\}$ and sub-band $q=11$ ($5600\,$MHz), using a sample size of $S_\text{K}=630$ samples at different time instances for a \emph{{general LOS obstruction on highway}} scenario.}%
\label{fig:04:12}
\end{center}
\end{figure}

Figure \ref{fig:04:12} (a), (b), and (c) show the CDF of the first five delay bins $n\in\{0,\ldots,4\}$ from $\mid h'[m,n;q]\mid$. The measured data is represented by a solid line and the fitted {CDF} uses a dashed line. Results are shown for frequency sub-band $q=11$, corresponding to a frequency of $5600\,$MHz, at time instances $0.1$\,s, $4.8$\,s, and $9.8$\,s respectively. At $0.1$\,s the {LOS} between {TX} and {RX} is obstructed, therefore the distribution of the first tap (denoted 'tap 0') is closer to being Rayleigh (note the shift towards the left hand side of the plot). At $4.8\,$s, the truck blocking the {LOS} has moved to another lane allowing a {LOS} communication. As a result, the $K$-factor of the first delay bin increases and its {CDF} is shifted to the right in Fig. \ref{fig:04:12} (b). The same effect can be noticed at $9.8\,$s (Fig. \ref{fig:04:12} (c)).

\begin{figure}
\begin{center}
\subfigure[$t=9.8\,$s]{\includegraphics[width=1\columnwidth]{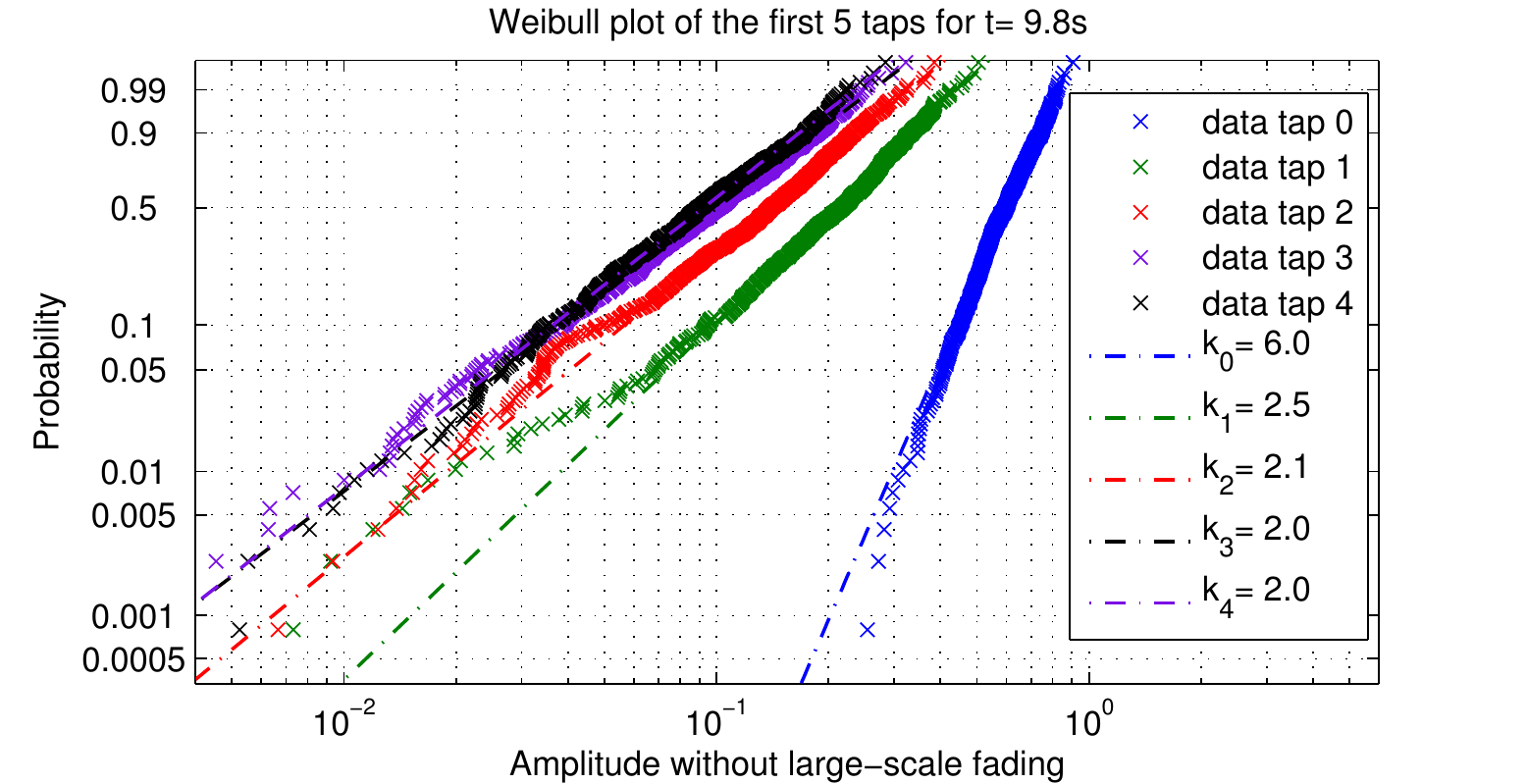}} 
\caption{Weibull plot for delay bins $n\in\{0,\ldots,4\}$  
and sub-band $q=11$ ($5600\,$MHz), using a sample size of $S_\text{K}=630$ samples for $t=9.8\,\text{s}$ for a \emph{{general LOS obstruction on highway}} scenario.}%
\label{fig:wblplot}
\end{center}
\end{figure} 

Furthermore, we observe that delay bins $n\in\{1, \ldots,4\}$ (denoted 'tap 1\dots4') follow a Rayleigh distribution throughout all time instances. On the other hand, the first delay bin is clearly Rician distributed with a varying $K$-factor for the three different time instances, $K_{t=0.1\,\text{s}}=5.77\,$dB, $K_{t=4.8\,\text{s}}=16.51\,$dB, and $K_{t=9.8\,\text{s}}=11.16\,$dB.

For $t=9.8\,\text{s}$ we also show an examplary Weibull plot \cite{McCool12} of the first 5 tap distributions in Fig. \ref{fig:wblplot}. The Weibull distribution 
\begin{equation}
f(z;\lambda,k)=\left\{\begin{array}{ll}
\frac{k}{\lambda}\left(\frac{z}{\lambda}\right)^{k-1}e^{-(z/\lambda)^k} & z\ge 0\,, \\
0 & z<0\,,\\
\end{array}\right.
\end{equation}
where $k>0$ is the shape parameter and $\lambda>0$ is the scale parameter. The CDF is given as
\begin{equation}
F(z;\lambda,k)=1-e^{-(z/\lambda)^k}\,,
\end{equation}
for $z\geq 0$, and $F(z; \lambda, k)=0$ for $z<0$. For $k=2$ the Weibull distribution is identical with the Rayleigh distribution and for $k > 2$ it is analogous to a Rician distribution \cite{Renaudin2010}, \cite{Matolak11}. For the Weibull plot the empirical cumulative distribution function $\hat{F}(z)$ is shown using axes $\ln(-\ln(1-\hat{F}(z)))$ vs. $\ln(z)$ resulting in a linearization of the Weibull CDF
\begin{equation}
\underbrace{\ln(-\ln(1-\hat{F}(z)))}_y=\underbrace{k\ln z}_{k x}-\underbrace{k\ln \lambda}_c.
\end{equation}
From Fig. \ref{fig:wblplot} we can see that the taps $n\in\{1,\ldots,4\}$ have a Rayleigh distribution with shape parameter $k_n\approx 2$ while tap $n=0$ is clearly Rician distributed with $k_0>2$.
We choose the time instance $t=9.8$\,s for exemplification because of the clear values of the shape parameter $k_n$, however, this is also valid for the other time instances, being their shape parameters listed in Tab. \ref{tab:Weibull}.
\begin{table}
\begin{center}
\caption{Shape parameters for the Weibull distribution}
\label{tab:Weibull}
\begin{tabular}{llllclc} 
\toprule
shape parameter             & $t=0.1$\,s	& $t=4.8$\,s	& $t=9.8$\,s  \\
\midrule
$k_0$            			& $3.4$		& $9.2$		& $6.0$ \\
$k_1$            			& $2.3$		& $3.5$		& $2.5$\\
$k_2$            			& $2.1$		& $2.1$		& $2.2$\\
$k_3$            			& $1.9$		& $2.0$		& $2.0$\\
$k_4$            			& $2.0$		& $2.1$		& $2.0$\\
\bottomrule
\end{tabular}
\end{center}
\end{table}

\tz{The analysis in this section for the the delay bins $n\in\{0,\ldots, 4\}$ clearly demonstrates, that the delay bin $n=0$ is strongly Rician distributed with a $K$ factor larger than zero while the delay bin $n\in\{1,\ldots, 4\}$ are mostly Rayleigh distributed. Hence, we will focus our analysis on delay bin $n=0$ in this paper.}

\subsection{$K$-Factor Estimation}
The  $K$-factor estimate 
\begin{equation}
\hat{K}[m,n;q]=F\Big\{h'[m',n;q]\, | \, m'\in\{m-\frac{S_\text{K}}{2},\ldots, m +\frac{S_\text{K}}{2}-1\}\Big\}
\end{equation}
is calculated for the first delay bins $n\in\{0,\ldots,4\}$, where $F\{\cdot\}$ denotes the $K$-factor estimation from a finite set of observation at time indices $\{m-\frac{S_\text{K}}{2},\ldots, m +\frac{S_\text{K}}{2}-1\}$. For estimating the $K$-factor we use a technique introduced in \cite{Oestges2010} which is based on the method of moments (MoM) \cite{Greenstein1999}.

\section{Empirical Results}
\label{sec:empiricalresults}
The goal of this investigation is to characterize the variation of the $K$-factor in time and frequency. The analysis is only carried out for the first delay bin $n=0$. 

\begin{figure*}
\begin{center}
\subfigure[Time-frequency dependent $K$-factor.]{\includegraphics[width=0.48\textwidth]{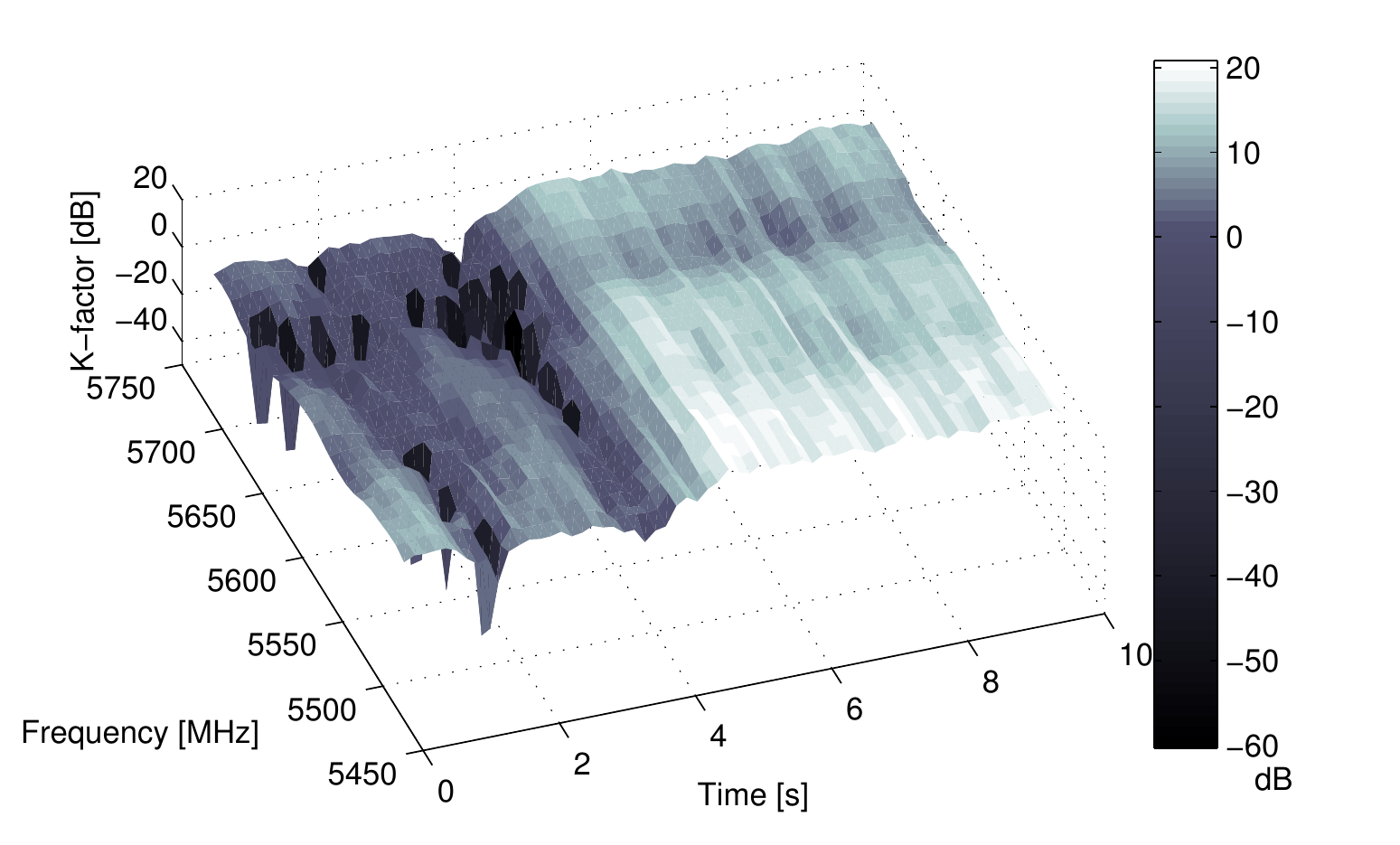}}
\subfigure[Time-frequency dependent power.]{\includegraphics[width=0.48\textwidth]{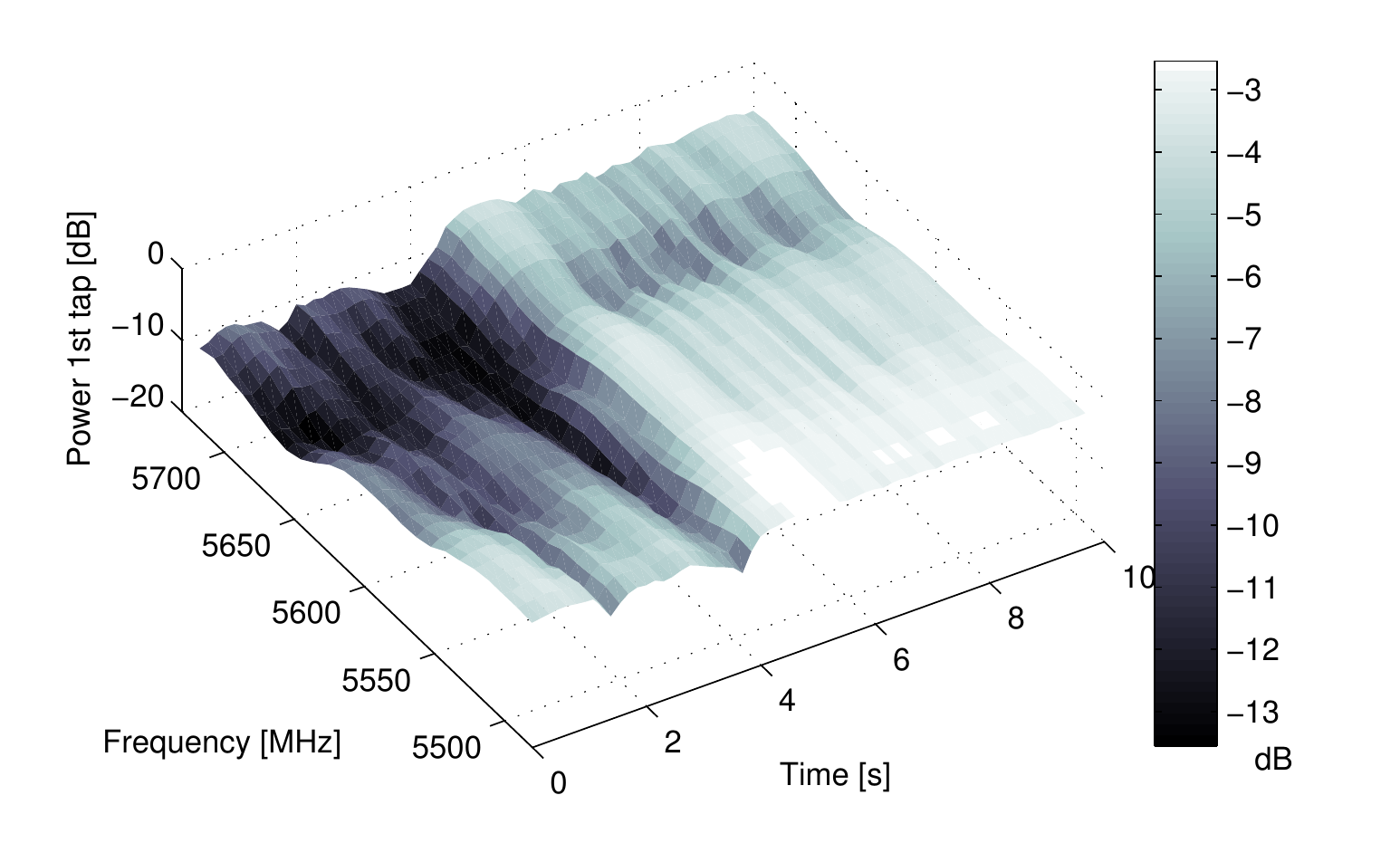}}
\caption{\emph{General LOS obstruction on highway}: Time-frequency dependent $K$-factor $\hat{K}[m,0;q]$ and power for tap $n=0$ of link $\ell=10$.}
\label{fig:04:13}
\end{center}
\end{figure*}

The time- and frequency-varying $K$-factor estimate $\hat{K}[m,0;q]$ is plotted in Fig. \ref{fig:04:13}, next to its corresponding time- and frequency varying power of the first delay bin. Here, the time-frequency variation of the $K$-factor can be appreciated already. 

Noteworthy is that there is not necessarily a correspondence between received power and $K$-factor (see the scale in the figures). This is due to the fact that the received power is calculated as the sum of the power of all specular and diffuse components, $P=r^2+2\sigma^2$, whereas the $K$-factor is an indicator of the ratio between them, see (\ref{eq:Kfactor}).

On what follows, we discuss the time and frequency variability of the $K$-factor of the first delay bin independently. We remind here that the chosen illustrative measurement corresponds to a \emph{{general LOS obstruction on highway}} scenario. In the studied measurement run, the LOS between the {TX} and the {RX} cars is obstructed for the first $5\,$s, afterwards, the truck blocking the {LOS} moves away.

\tz{Due to the large variation of the $K$ factor all following figure displaying the $K$-factor will be given in logarithmic scale. We would like to point out, that for regions where the $K$-factor is close to $0$, the logarithmic scale might overemphasize its variation since the range $0 \ldots 1$ in linear scale is mapped to the range $-\infty \ldots 0$ in logarithmic scale. Please keep this in mind when interpreting the figures.}

\subsection{Time-Varying $K$-Factor}
\begin{figure}
\begin{center}
{\includegraphics[width=0.90\columnwidth]{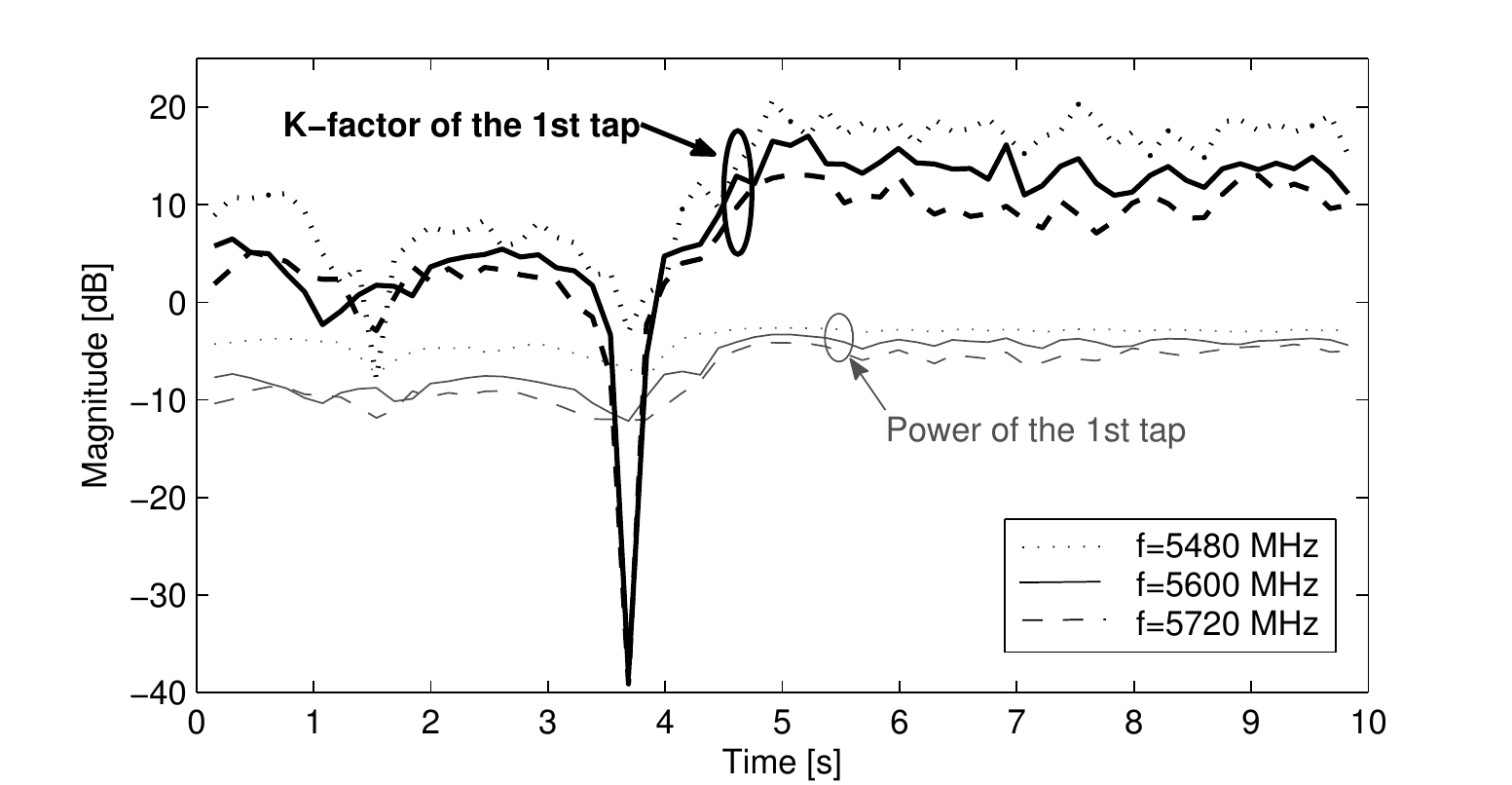}}
\caption{\emph{General LOS obstruction on highway}: Time-varying $K$-factor and power of the first delay tap $n=0$ without large-scale fading for three different frequency sub-bands $f=q f_\text{s}\in\{5.48, 5.6, 5.72\}\,\text{GHz}$ for a sample size of $S_\text{K}=630$.}
\label{fig:04:16}
\end{center}
\end{figure}

Figure \ref{fig:04:16} shows the time evolution of the $K$-factor for three different frequency sub-bands, together with their corresponding power. The three curves are obtained by making a cut in Fig. \ref{fig:04:13} at the three different frequency sub-bands. The $K$-factors obtained at the three frequency bands are comparable although they do not have the same numerical value. Furthermore, they change in time following a similar tendency.

Large $K$-factors up to $20\,$dB are observed between $5$ and $10$\,s. However, the estimated $K$-factor is around $10\,$dB smaller for the period from $0$ to $5$\,s. In this case, there are more significant diffuse components in the received signal, and thus it is better described by a lower $K$-factor. We call obstructed LOS (oLOS) the situation when the LOS between TX and RX is obstructed, but the waves still propagate by diffraction over the roof-top of the obstructing vehicle, as it occurs between seconds $0$ and $5$.

We observe estimated $K$-factor values for the oLOS to be between $0$ and $5$\,dB, being those lower than the $K$-factor in pure LOS communication, with values around $15$\,dB. Lower $K$-factors are obtained in non-LOS (nLOS) situations (shown in results in Sec. \ref{sec:modeling}). Thus, we can establish the following relationship: $K_\text{nLOS}<<K_\text{oLOS}<K_\text{LOS}$.

\subsection{Frequency-Varying $K$-Factor}
\label{subsec:K-frequency}
\begin{figure}
\begin{center}
\includegraphics[width=0.90\columnwidth]{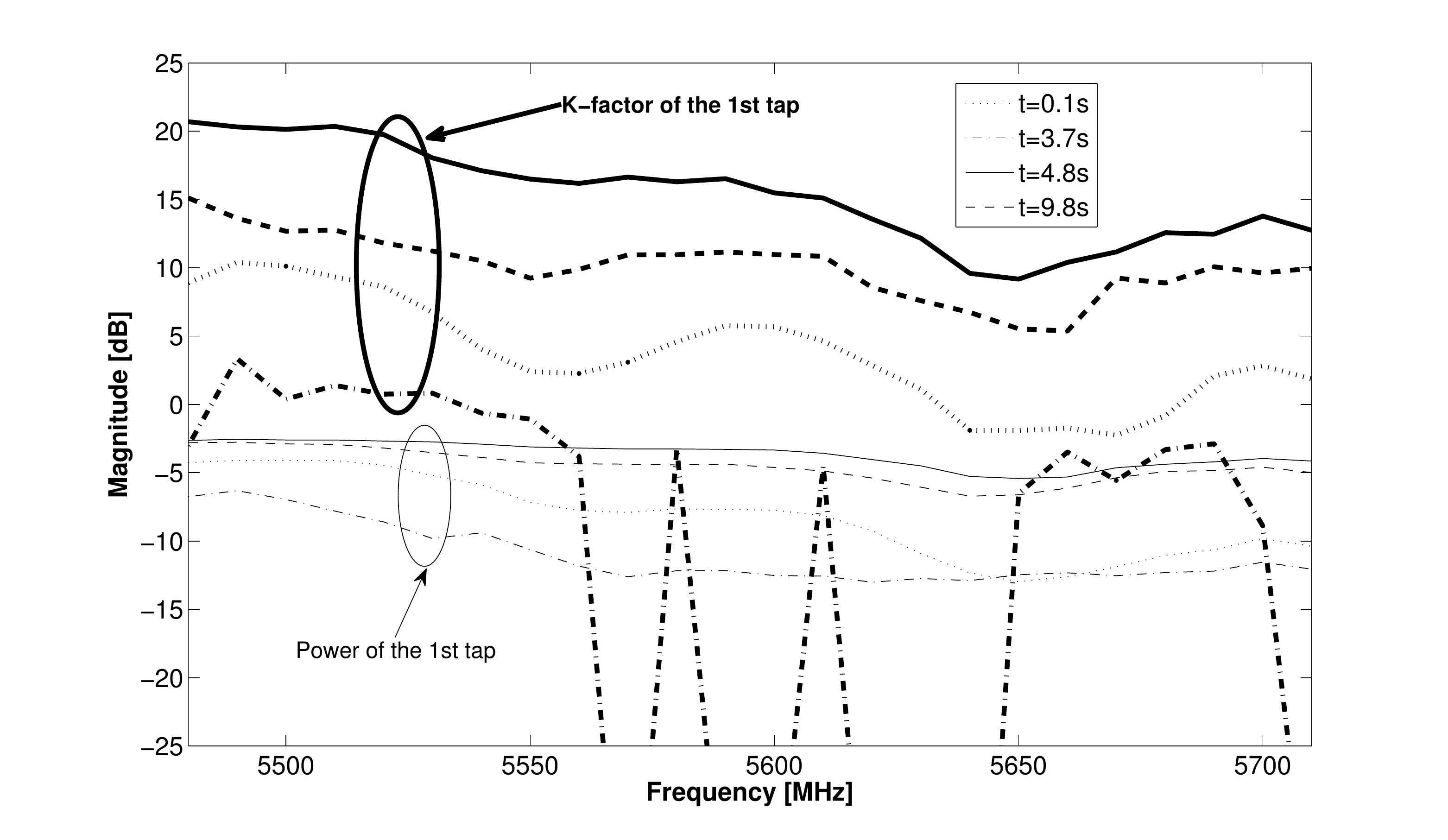}
\caption{\emph{General LOS obstruction on highway}: Frequency-varying $K$-factor and power of the first delay tap $n=0$ without large-scale fading for three different time instances \tz{$t=m t_\text{s}\in\{0.1, 3.7, 4.8, 9.8\}\,\text{s}$} for a sample size of $S_\text{K}=630$.}
\label{fig:04:17}
\end{center}
\end{figure}

For the frequency-variation analysis we select \tz{four} different time instances and plot their estimated $K$-factor as a function of $Q=24$ frequency sub-bands in Fig. \ref{fig:04:17}, together with their corresponding power. The three curves here are obtained by making a cut in Fig. \ref{fig:04:13} at three different time instances $t=m t_\text{s}\in\{0.1, 3.7, 4.8, 9.8\}\,\text{s}$. Looking at the figure, it can be concluded that the assumption of a constant $K$-factor in frequency (within the $240\,$MHz measured frequency band) is not realistic.

Variations of up to $12\,$dB are observed throughout the $24$ frequency sub-bands, mainly due to the frequency dependent antenna radiation pattern. The variation of the antenna gains for antenna elements $n_\text{TX}=3$ and $n_\text{RX}=3$ shown in Fig. \ref{fig:04:15} are significant, when comparing the gains at the lower, central, and upper frequency band. This effect is mostly observed for element $n_\text{RX}=3$, where the gain experiences a variation of $10$\,dB at the bandwidth edges with respect to the carrier frequency of $5600\,$MHz. The directionality of the pattern is affected by the frequency variations, but notches are significant, such as at $50^\circ$, and $280^\circ$, giving rise to drops at some frequencies, as observed in Fig. \ref{fig:04:13} (a). The $K$-factor also varies in frequency due to the effect of foliage and vegetation in the surrounding environment.

\tz{The results for $t=3.7$ are shown in Fig. \ref{fig:04:17} to provide a link to Fig. \ref{fig:04:16} and Fig. \ref{fig:04:13}.a where a very low $K$-factor near zero, or in the logarithmic domain of $-\infty$, is obtained for this time instant.}

\section{Statistical Modeling}
\label{sec:modeling}
\subsection{Distribution Fitting}
We are interested in characterizing the distribution of the $K$-factor for the sub-band bandwidth of $10\,\text{MHz}$. In order to see whether we can use the whole data ensemble of the link for the statistical characterization, we first have a look at the {CDF} of the envelope of the first tap for the $Q=24$ frequency sub-bands individually, as done in \tz{Sec. \ref{se:EvelopeDistributionTap}} for a single sub-band. The result of applying the KS-test shows values of $\epsilon < 0.04$ for all sub-bands, which we consider low enough and therefore conclude that the first tap is described by a Rician distribution in all $Q$ sub-bands. Therefore we will use the estimated $K$-factors from all available $Q$ sub-bands jointly for fitting a distribution.

\begin{figure}
\begin{center}
\includegraphics[width=\columnwidth]{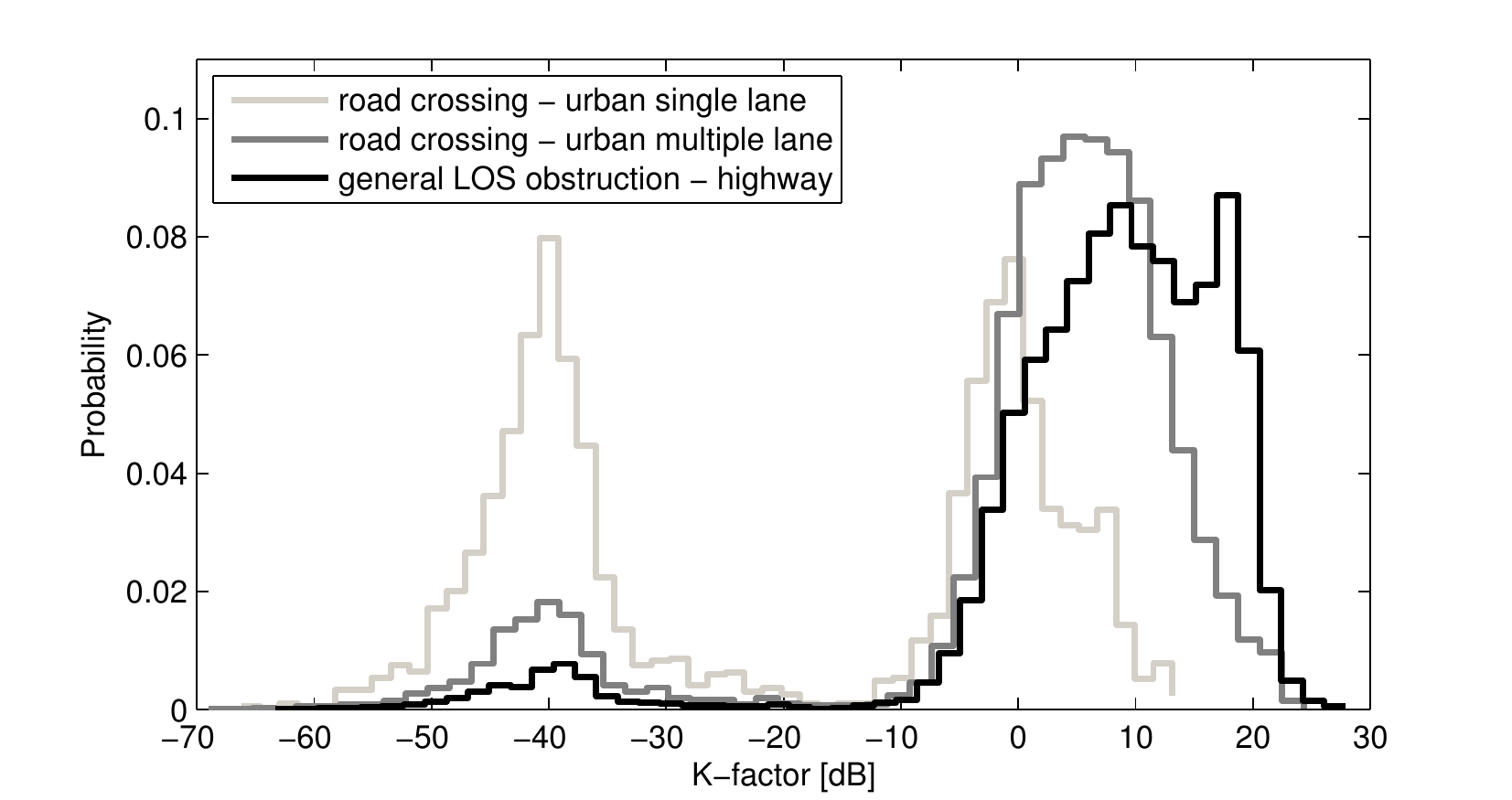}
\caption{Normalized histogram (histogram/number of samples) for three different scenarios.}
\label{fig:pmf}
\end{center}
\end{figure}

We show in Fig. \ref{fig:pmf} the normalized histogram (histogram divided by the total number of samples) for three different scenarios. Two Gaussian-shaped distributions can be envisioned from these histograms, which motivates us to select a bi-modal Gaussian distribution for fitting purposes, similar as in the companion paper for the root-mean-square (RMS) Delay and Doppler Spread \cite{Bernado2014}.

One could argue that the pdf of the K-factor is different depending on the selected sample ensemble, which is time-dependent, for each measurement run. For instance, for the exemplary \emph{general LOS obstruction} scenario, if we separate the CIRs of our measurement runs into two ensembles, one set containing the CIRs when the LOS is not obstructed and the other one when the LOS is obstructed, the resulting pdf of the K-factor for the two sets of CIRs will be different. However, we consider adequate to characterize both sets jointly with a single bi-modal distribution, in order to describe the transition probability between them as well.

The probability density function (pdf) $p$ of a bi-modal Gaussian mixture distribution reads
\begin{equation}
p(K)=\frac{w}{\sqrt{2\pi}\,\sigma_1}e^{-\frac{(K-\mu_1)^2}{2\sigma_1}} + \frac{1-w}{\sqrt{2\pi}\,\sigma_2}e^{-\frac{(K-\mu_2)^2}{2\sigma_2}},
\label{eq:BiModalGaussian}
\end{equation}
where $\mu_x$ and $\sigma_x$ are the mean and the standard deviation of each Gaussian component with index $x\in\{1,2\}$, and $w$ and $1-w$ are the weighting factors of the first and second component respectively. The first Gaussian function describes the nLOS situation (small $K$-factor), the other one describes the {LOS} (large $K$-factors) situation. The weighting factors of the two Gaussians $w$ and $1-w$ indicate the probability of having nLOS or {LOS} for a particular scenario. 

\begin{figure}
\begin{center}
\includegraphics[width=\columnwidth]{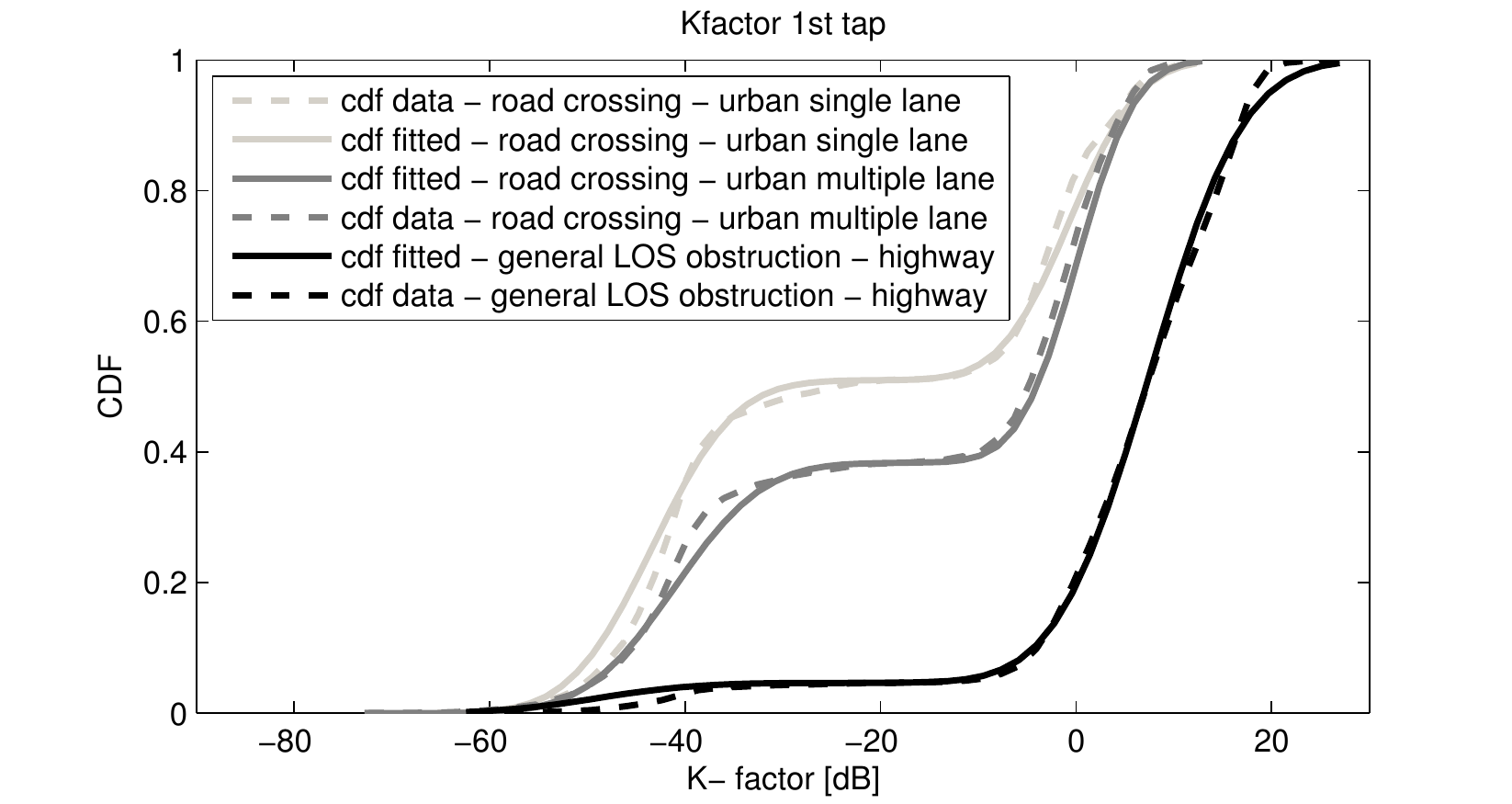}
\caption{\emph{General LOS obstruction on highway}: Joint {CDF} - time-frequency dependent $K$-factor.}
\label{fig:cdf}
\end{center}
\end{figure}

\tz{In Fig. \ref{fig:cdf}, we plot the empirical {CDF} $P_K(K)$ of the estimated $K$-factor as solid line, and the fitted {CDF} 
\begin{equation}
P_0(K)= w\left(1 - Q\left(\frac{K-\mu_1}{\sigma_1}\right) \right) + (1-w)\left(1 - Q\left(\frac{K-\mu_2}{\sigma_2}\right) \right)
\end{equation}
as dashed line.

We apply the KS test as a GoF indicator
\begin{equation}
\epsilon =\sup_z|P_K(K)-P_0(K)|
\end{equation} 
to assess the accuracy of the bi-modal Gaussian distribution \cite{Bernado2014}. For all safety relevant measurement scenarios we obtain $\epsilon<0.06$, thus showing a very good fit, as listed in Tab. \ref{tab:04:08}. Hence, the bi-modal Gaussian distribution (\ref{eq:BiModalGaussian}), based on only five parameters, is a simple, but at the same time accurate model for the $K$-factor distribution in safety-relevant vehicular scenarios.}

\subsection{Discussion}
\subsubsection{Weighting factor $w$}
Looking at the results, in dB, in Tab. \ref{tab:04:08}, we see that in some cases the weighting factor of the one of the component is very small, namely for the \emph{general LOS obstruction, merging lanes}, and \emph{approaching traffic jam}, where one of the weighting factors is below $0.10$. 
Comparable weighting factors for the both components of the bi-modal Gaussian distribution are obtained for \emph{road crossing - urban} scenarios, for both \emph{single lane} and \emph{multiple lanes}. Precisely in these two scenarios the cars experience nLOS conditions more often, since there are buildings in the four quadrants of the crossing. 

A particular case exhibiting similar weighting factors is the \emph{on-bridge} scenario. Nevertheless, here we do not have two clear Gaussian distributions, since the two mean values, $\mu_1$ and $\mu_2$ are less than $4$\,dB apart from each other. This indicates that the actual shape of the histogram for this particular scenario is close to a single Gaussian.

The other scenarios, \emph{road crossing in suburban areas, slow traffic,} and \emph{in-tunnel}, have different weighting factors, but with very dis-similar mean values for each component.

\subsubsection{Mean value $\mu$}

In general, the two mean components, $\mu_1$ and $\mu_2$ are well separated being $\mu_1<\mu_2$. The lower component $\mu_1$ corresponds to the nLOS period of the measurement, as discussed previously, with values around $-40$\,dB in most of the cases (we have already commented on the special case of \emph{on-bridge} scenario).

If we now concentrate on $\mu_2$, the highest value is obtained for the \emph{merging lanes} (14.2\,dB) {and \emph{on-bridge} (14.6\,dB)} scenarios, where there is a clear LOS (the two cars drive one after the other without obstacles between them).

On the other extreme, the lowest $\mu_2$ are obtained in the two scenarios in \emph{urban} environment, \emph{road crossing - single lane} (-0.6\,dB) and \emph{road crossing - multiple lane} (0.11\,dB), with comparable values. In these measurements (\emph{urban}) the number of scatterers close to the TX-RX link is large, leading to low $K$-factors \tz{\cite{Abbas2011}}.

In general, scatterers are closer in \emph{road crossing} scenarios. Nevertheless, in \emph{suburban} environments, their number is lower, and $\mu_2$ for the \emph{road crossing - suburban} scenarios are a bit higher than in \emph{urban} environments. The mean value of the $K$-factor, $\mu_2$, for situations \emph{with traffic} (3.7\,dB) is sightly lower than $\mu_2$ for situations \emph{without traffic} (4.5\,dB), as the more traffic (i.e. more vehicles around), the more scattering. The rich scattering leading to low $\mu_2$ values is also present in \emph{in-tunnel} ($4.7\,$dB), and \emph{slow traffic} ($4.4\,$dB) scenarios. Finally, the $\mu_2$ values for the two scenarios on the \emph{highway} are also comparable, being 7.6\,dB for the \emph{general LOS obstruction} scenario, and 8.1\,dB for the \emph{approaching traffic jam} scenario.

To summarize, we can list the $\mu_2$ values from low to high and also depending on the richness of the scattering environment as: \emph{crossing-urban} (-0.6, 0.1\,dB) $<$ \emph{crossing-suburban} (3.7, 4.5\,dB) $<$ \emph{slow-traffic} (4.4\,dB) $\approx$ \emph{in-tunnel} (4.7\,dB) $<$ \emph{highway (general LOS obstruction and approaching traffic jam)} (7.6, 8.1\,dB) $<$ \emph{merging lanes} (14.2\,dB).

\subsubsection{Standard deviation $\sigma$}

Nearly all $\sigma_1$ have comparable values among themselves, as well as all $\sigma_2$ among themselves. The standard deviation of the first component $\sigma_1$ is around 7\,dB, except for the \emph{merging lane} scenario, which is 21.7\,dB. We do not give importance to this value, since the the first component in this particular scenario is almost negligible ($w=0.03$).

The standard deviation of the second component, $\sigma_2$, is around 5\,dB for all scenarios, except for the \emph{general LOS obstruction} (7.5\,dB) and the two \emph{traffic congestion} (6.5\,dB) scenarios. In these two cases, oLOS occurs with higher frequency, with values close to the LOS case (discussed in Sec. \ref{sec:empiricalresults}), thus being reflected in larger standard deviations in our modeling.

\begin{table}
\caption{Modeling the time-varying $K$-factor}
\label{tab:04:08}
\begin{center}
 \begin{tabular}{@{}rrrrrrr@{}}\toprule
$w$ & $\mu_1$\,[dB] & $\sigma_1$\,[dB] & $\mu_2$\,[dB] & $\sigma_2$\,[dB] & $\epsilon$ & runs\\
\midrule
\multicolumn{7}{@{}l@{}}{\emph{\textbf{road crossing - suburban with traffic}}}\\
$0.27$ & $-42.7$ & $7.5$  & $3.7$ & $5.2$  & $<0.04$  & $3$\\\addlinespace[1mm]
\midrule
\multicolumn{7}{@{}l@{}}{\emph{\textbf{road crossing - suburban without traffic}}}\\
$0.13$  & $-43.0$ & $7.7$ & $4.5$ & $5.5$ & $<0.02$ & $11$\\\addlinespace[1mm]
\midrule
\multicolumn{7}{@{}l@{}}{\emph{\textbf{road crossing - urban single lane}}}\\
$0.51$   & $-43.3$ & $6.6$  & $-0.6$ & $5.6$  & $<0.06$ & $5$\\\addlinespace[1mm]
\midrule
\multicolumn{7}{@{}l@{}}{\emph{\textbf{road crossing - urban multiple lane}}}\\
$0.38$   & $-41.1$ & $7.2$   & $0.1$ & $4.7$  & $<0.05$ & $5$\\\addlinespace[1mm]
\midrule
\multicolumn{7}{@{}l@{}}{\emph{\textbf{general LOS obstruction - highway}}}\\
$0.05$   & $-48.9$ & $7.9$& $7.6$ & $7.5$  & $<0.04$ & 12\\\addlinespace[1mm]
\midrule
\multicolumn{7}{@{}l@{}}{\emph{\textbf{merging lanes - rural}}}\\
$0.03$   & $-29.9$ & $21.7$& $14.2$ & $4.2$  & $<0.02$ & $7$\\\addlinespace[1mm]
\midrule
\multicolumn{7}{@{}l@{}}{\emph{\textbf{traffic congestion - slow traffic}}}\\
$0.12$   & $-43.1$ & $8$ & $4.4$ & $6.5$ & $<0.02$ & 11\\\addlinespace[1mm]
\midrule
\multicolumn{7}{@{}l@{}}{\emph{\textbf{traffic congestion - approaching traffic jam}}}\\
$0.03$   & $-49.2$ & $7.9$ & $8.1$ & $6.5$ & $<0.02$ & $7$\\\addlinespace[1mm]
\midrule
\multicolumn{7}{@{}l@{}}{\emph{\textbf{in-tunnel}}}\\
$0.10$   & $-43.1$ & $7.2$ & $4.7$ & $5.4$ & $<0.02$ & $7$\\\addlinespace[1mm]
\midrule
\multicolumn{7}{@{}l@{}}{\emph{\textbf{{on-bridge}}}}\\
$0.44$& {$10.9$}& {$3.2$}& {$14.6$}& {$4.2$} & {$<0.02$} & {$4$}\\
\bottomrule
\end{tabular}
\end{center}
\end{table}  

\section{Conclusions}
\label{sec:conclusions}
In this contribution, we analyzed and characterized the small scale fading from vehicular channel measurements. We found that the first delay bin is Rician distributed with a varying $K$-factor, and the following delay bins are mostly Rayleigh distributed. We also showed that the $K$-factor varies in the time- and frequency-domain. These variations are significant and must be taken into account for modeling the propagation conditions for safety relevant scenarios in intelligent transportation systems (ITS). 

The $K$-factor variation depends mainly on the following factors: \emph{(i)} The number of illuminated scatterers due to the antenna radiation pattern and objects in between TX and RX, \emph{(ii)} the antenna radiation pattern variation over the measurement bandwidth, \emph{(iii) vegetation and foliage}, and \emph{(iv)} the presence of good reflecting objects in the vicinity of the TX-RX link (e.g. traffic signs). 

Furthermore, we provided a statistical model for the $K$-factor of the first delay bin in a $10\,$MHz communication bandwidth for important safety critical vehicular scenarios matched to the IEEE 802.11p standard. For that, we used a bi-modal Gaussian distribution, where each one of the components models the $K$-factor in LOS and nLOS propagation conditions.

The weighting factors can be interpreted as the transition probabilities between LOS and nLOS situations. They are similar for \emph{road-crossing} scenarios in urban environments. Otherwise, one component is more important than the other (with a higher weighting factor), except for the \emph{general LOS obstruction, merging lanes}, and \emph{traffic congestion} scenarios, where the obstructed LOS (oLOS) situation is more common than nLOS. 

We also saw that the mean values corresponding to the nLOS Gaussian component are around $-40\,$dB. The mean values corresponding to LOS and oLOS situations are different depending on the scattering richness of the surroundings, being $-0.6$\,dB the lowest for \emph{road-crossing} in urban environment, and $14.2\,$dB for \emph{merging lanes} in rural environment.

Finally, we also analyzed the standard deviation of the two Gaussian components showing that the largest values are obtained for situations with LOS and oLOS mixed, since their mean values are close enough.
 
We consider that the work presented in this manuscript is the basis for more complex modeling of a variable $K$-factor, where time correlation can be included. Furthermore, the bi-modal Gaussian fitting can open the doors to a Markov-chain model of the $K$-factor for non-stationary channels.

\bibliography{IEEEabrv,biblio_diss}

\end{document}